\documentclass[letterpaper,11pt]{article}
\pdfoutput=1

\usepackage{jheppub}
\usepackage{multirow}
\usepackage{braket}
\usepackage{mathrsfs}
\usepackage{tikz}
\usepackage{subfig}
\usepackage{xspace}
\usepackage{xcolor}
\usepackage{float,color}
\usepackage{siunitx}
\usepackage{comment}
\usepackage[utf8]{inputenc}
\usepackage[countmax]{subfloat}

\providecommand{\href}[2]{#2}

\definecolor{darkred}{rgb}{0.5,0.0,0.0}
\definecolor{darkblue}{rgb}{0.0,0.0,0.9}
\definecolor{darkerblue}{rgb}{0.0,0.0,0.5}
\definecolor{darkgreen}{rgb}{0.0,0.5,0.0}
\definecolor{black}{rgb}{0.0,0.0,0.0}
\definecolor{brown}{rgb}{0.6,0.4,0.2}

\title{\boldmath Pileup Mitigation with Machine Learning (PUMML)}
 
\preprint{ 
\begin{flushright}
MIT--CTP 4924
 \end{flushright}}

\author[a]{Patrick T. Komiske,}
\author[a]{Eric M. Metodiev,}
\author[b]{Benjamin Nachman,}
\author[c]{Matthew D. Schwartz}

\affiliation[a]{Center for Theoretical Physics, Massachusetts Institute of Technology, Cambridge, MA 02139, USA}
\affiliation[b]{Physics Division, Lawrence Berkeley National Laboratory, Berkeley, CA 94720, USA}
\affiliation[c]{Department of Physics, Harvard University, Cambridge, MA 02138, USA}

\emailAdd{pkomiske@mit.edu}
\emailAdd{metodiev@mit.edu}
\emailAdd{bpnachman@lbl.gov}
\emailAdd{schwartz@physics.harvard.edu}

\abstract{
Pileup involves the contamination of the energy distribution arising from the primary collision of interest (leading vertex) by radiation from soft collisions (pileup). We develop a new technique for removing this contamination using machine learning and convolutional neural networks. The network takes as input  the energy distribution of charged leading vertex particles, charged pileup particles, and all neutral particles and outputs the energy distribution of particles coming from leading vertex alone. The PUMML algorithm performs
remarkably well at eliminating pileup distortion on a wide range of simple and complex jet observables. We test the robustness of the algorithm in a number of ways and discuss how the network can be trained directly on data.
}

\begin{document} 
\maketitle
\flushbottom

\section{Introduction}
The Large Hadron Collider (LHC) is operated at very high instantaneous luminosities to achieve the large statistics required to search for exotic Standard Model (SM) or beyond the SM processes as well as for precision SM measurements. At a hadron collider, protons are grouped together in bunches; as the luminosity increases for a fixed bunch spacing, the number of protons within each bunch that collide inelastically increases as well. Most of these inelastic collisions are soft, with the protons dissolving into mostly low-energy pions that disperse throughout the detector. A typical collision of this sort at the LHC will contribute about 0.6 GeV/rad${}^2$ of energy~\cite{Khachatryan:2016kdb,Aaboud:2017jcu}. Occasionally, one pair of protons within a bunch crossing collides head-on, producing hard (high-energy) radiation of interest. At high luminosity, this hard collision, or leading vertex (LV), is always accompanied by soft proton-proton collisions called pileup. The data collected thus far by ATLAS and CMS have approximately 20 pileup collisions per bunch crossing on average ($\langle\text{NPU}\rangle \sim 20$); the data in Run 3 are expected to contain $\langle\text{NPU}\rangle\sim 80$; and the HL-LHC in Runs 4-5 will have $\langle\text{NPU}\rangle\sim 200$. Mitigating the impact of this extra energy on physical observables is one of the biggest challenges for data analysis at the LHC.

Using precision measurements, the charged particles coming from the pileup interactions can mostly be traced to collision points (primary vertices) different from that of the leading vertex. Indeed, due the to the excellent vertex resolution at ATLAS and CMS~\cite{Chatrchyan:2014fea,ATLAS-CONF-2010-027,ATLAS-CONF-2010-069} the charged particle tracks from pileup can almost completely be identified and removed.\footnote{Some detector systems have an integration time that is (much) longer than the bunch spacing of $\SI{25}{ns}$, so there is also a contribution from pileup collisions happening before or after the collision of interest (out-of-time pileup).  This contribution will not have charged particle tracks and can be at least partially mitigated with calorimeter timing information. Out-of-time pileup is not considered further in this analysis.} This is the simplest pileup removal technique, called {\it charged-hadron subtraction}. The challenge with pileup removal is therefore how to distinguish neutral radiation associated with the hard collision from neutral pileup radiation.\footnote{Charged-hadron subtraction follows a particle-flow technique that removes calorimeter energy from pileup tracks.  Due to the calorimeter energy resolution, there will be a residual contribution from charged-hadron pileup.  This contribution is ignored but could in principle be added to the neutral pileup contribution.} Since radiation from pileup is fairly uniform\footnote{This work will not explicitly discuss identification of real high energy jets resulting from pileup collisions.  The ATLAS and CMS pileup jet identification techniques are documented in Ref.~\cite{Aad:2015ina,Aaboud:2017pou} and~\cite{CMS:2013wea}, respectively. }, it can be removed on average, for example, using the  jet areas technique~\cite{Cacciari:2007fd}. The jet areas technique focuses on correcting the overall energy of collimated sprays of particles known as jets. Indeed, both the ATLAS and CMS experiments apply jet areas or similar techniques to calibrate the energy of their jets~\cite{Chatrchyan:2011ds,Aad:2011he,Khachatryan:2016kdb,ATLAS-CONF-2015-037,CMS-DP-2016-020,Aaboud:2017jcu}. Unfortunately, for many measurements, such as those involving jet substructure or the full radiation patterns within the jet, removing the radiation on average is not enough.  

Rather than calibrating only the energy or net 4-momentum of a jet, it is possible to correct the constituents of the jet.  By removing the pileup contamination from each constituent, it should be possible to reconstruct more subtle jet observables. We can coarsely classify constituent pileup mitigation strategies into several categories: constituent preprocessing, jet/event grooming, subjet corrections, and constituent corrections.  Grooming refers to algorithms that remove objects and corrections describe scale factors applied to individual objects. Both ATLAS and CMS perform preprocessing to all of their constituents before jet clustering.  For ATLAS, pileup-dependent noise thresholds in topoclustering~\cite{Aad:2016upy} suppresses low energy calorimeter deposits that are characteristic of pileup.  In CMS, charged-hadron subtraction removes all of the pileup particle-flow candidates~\cite{Sirunyan:2017ulk}.  Jet grooming techniques are not necessarily designed to exclusively mitigate pileup but since they remove constituents or subjets in a jet (or event) that are soft and/or at wide angles to the jet axis, pileup particles are preferentially removed~\cite{Butterworth:2008iy, Krohn:2009th, 0903.5081, 0912.0033, Larkoski:2014wba,Cacciari:2014gra,Aad:2015ina}.  Explicitly tagging and removing pileup subjets often performs comparably to algorithms without explicit pileup subjet removal~\cite{Aad:2015ina}.  A popular event-level grooming algorithm called SoftKiller~\cite{Cacciari:2014gra} removes radiation below some cutoff on transverse momentum, $p_T^{\text{cut}}$ chosen on an event-by-event basis so that half of a set of pileup-only patches are radiation free.  

While grooming algorithms remove constituents and subjets, there are also techniques that try to reconstruct the exact energy distribution from the primary collision. One of the first such methods introduced was Jet Cleansing~\cite{Krohn:2013lba}. Cleansing works at the subjet level, clustering and declustering jets to correct each subjet separately based on its local energy information. Furthermore, Cleansing  exploits the fact that the relative size of pileup fluctuations decreases as $\langle\text{NPU}\rangle \to \infty$ so that the neutral pileup-energy content of subjets can be estimated from the charged pileup-energy content.  A series of related techniques operate on the constituents themselves~\cite{Bertolini:2014bba,Berta:2014eza,atlascvf}.  One such technique called PUPPI also uses local charged track information but works at the particle level rather than subjet level. PUPPI computes a scale factor for each particle, using a local estimate inspired by the jets-without-jets paradigm~\cite{Bertolini:2013iqa}. In this paper, we will be comparing our method to PUPPI and SoftKiller. 

In this paper, we present a new approach to pileup removal based on machine learning. The basic idea is to view the energy distribution of particles as the intensity of pixels in an image~\cite{Cogan:2014oua}.  Convolutional neural networks applied to jet images~\cite{deOliveira:2015xxd} have found widespread applications in both classification~\cite{deOliveira:2015xxd,Baldi:2016fql,Barnard:2016qma,Kasieczka:2017nvn,komiske2017} and generation~\cite{deOliveira:2017pjk,Paganini:2017hrr}. Previous jet-images applications have included boosted $W$-boson tagging~\cite{deOliveira:2015xxd,Baldi:2016fql,Barnard:2016qma}, boosted top quark identification~\cite{Kasieczka:2017nvn}, and quark/gluon jet discrimination~\cite{komiske2017}.  Most of these previous applications were all \emph{classification} tasks: extracting a single binary classifier (quark or gluon, $W$ jet or background jet, etc.) from a highly-correlated multidimensional input. The application to pileup removal is a more complicated \emph{regression} task, as the output (a cleaned-up image) should be of similar dimensionality to the input. PUMML is among the first applications of modern machine learning tools to regression problems in high energy physics.

To apply the convolutional neural network paradigm to cleaning an image itself, we exploit the finer angular resolution of the tracking detectors relative to the calorimeters of ATLAS and CMS. Building on the use of multichannel inputs in~\cite{komiske2017}, we give as input to our network three-channel jet images: one channel for the charged LV particles, one channel for the charged pileup particles, and one channel, at slightly lower resolution, for the total neutral particles. We then ask the network to reconstruct the unknown image for LV neutral particles. Thus our inputs are like those of Jet Cleansing but binned into a regular grid (as images) rather than single numbers for each subjet~\cite{Krohn:2013lba}. Further, the architecture is designed to be local (as with Cleansing or PUPPI), with the correction of a pixel only using information in a region around it. The details of our network architecture are described in Section~\ref{sec:algorithm}. Section~\ref{sec:performance} documents its performance in comparison to other state-of-the-art techniques. The remainder of the paper contains some robustness checks and a discussion in Section~\ref{sec:conc} of the challenges and opportunities for this approach. 

\section{PUMML algorithm}\label{sec:algorithm}

The goal of the PUMML algorithm is to reconstruct the neutral leading vertex radiation from the charged leading vertex, charged pileup, and total neutral information. Since neutral particles do not have tracking information available, the challenge is to determine what fraction of the total neutral energy in each direction came from the leading vertex and what fraction came from pileup. To assist this discrimination, we take as inputs into our network the energy distribution of charged particles, separated into leading vertex and pileup contributions, in addition to the total neutral energy distribution\footnote{Both ATLAS~\cite{CERN-LHCC-2015-020} and CMS~\cite{Butler:2055167,Contardo:2020886} are proposing precision timing detectors are part of their upgrades for the HL-LHC; such information could naturally be incorporated into another layer of the network.}. A natural way to combine these observables is using the multichannel images approach introduced in~\cite{komiske2017} based on color-image recognition technology. 

We apply this machine learning technique to $R=0.4$ anti-$k_t$ jets. The jet image inputs are square grids in pseudorapidity-azimuth $(\eta,\phi)$ space of size $0.9 \times 0.9$ centered on the charged leading vertex transverse momentum ($p_T$)-weighted centroid of the jet.  One could combine all layers to determine the jet axis, but in practice the axis determined from the charged leading vertex captures dominates because of its superior angular resolution and pileup robustness. To simulate the detector resolutions of charged and neutral calorimeters, charged images are discretized into $\Delta\eta\times\Delta\phi = 0.025\times 0.025$ pixels and neutral images are discretized into $\Delta\eta\times\Delta\phi = 0.1\times 0.1$ pixels\footnote{These dimensions are representative of typical tracking and calorimeter resolutions, but would be adapted to the particular detector in practice.  We ignore other detector effects in this algorithm demonstration, as has also been done also for PUPPI and SoftKiller.  In principle, additional complications due to the detector response can be naturally incorporated into the algorithm during training.}. We use the following three input channels:
\begin{enumerate}
\item[] {\color{darkred}{\tt red} = the transverse momenta of all neutral particles}
\item[] {\color{darkgreen}{\tt green} =  the transverse momenta of charged pileup particles}
\item[] {\color{darkblue}{\tt blue} = transverse momenta of charged leading vertex particles}
\end{enumerate}
The output of our network is also an image:
\begin{enumerate}
\item[] {\color{black}{\tt output} = the transverse momenta of neutral leading vertex particles.}
\end{enumerate}

Only charged particles with $p_T > 500$ MeV were included in the green or blue channels. Charged particles not passing this charged reconstruction cut were treated as if they were neutral particles. Otherwise, the separation into channels is assumed perfect. No image normalization or standardization was applied to the jet images, allowing the network to make use of the overall transverse momentum scale in each pixel. The different resolutions for charged and neutral particles initially present a challenge, since standard architectures assume identical resolution for each color channel. To avoid this issue, we perform a direct upsampling of each neutral pixel to $4\times 4$ pixels of size $\Delta\eta\times\Delta\phi = 0.025\times 0.025$ and divide each pixel value by 16 such that the total momentum in the image is unchanged.

In summary, the following processing was applied to produce the pileup images:
\begin{enumerate}
\item \emph{Center}: Center the jet image by translating in ($\eta,\phi$) so that the total charged leading vertex $p_T$-weighted centroid pixel is at $(\eta,\phi) = (0,0)$. This operation corresponds to rotating and boosting along the beam direction to center the jet.
\item \emph{Pixelate}: Crop to a $0.9\times 0.9$ region centered at $(\eta,\phi) = (0,0)$. Create jet images from the transverse momenta of all neutral particles, the charged leading vertex particles, the charged pileup particles, and the neutral leading vertex particles. Pixelizations of $\Delta\eta\times\Delta\phi = 0.025\times 0.025$ and $\Delta\eta\times\Delta\phi = 0.1\times 0.1$ are used for the charged and neutral jet images, respectively.
\item \emph{Upsample}: Upsample each neutral pixel to sixteen $\Delta\eta\times\Delta\phi = 0.025\times 0.025$ pixels, keeping the total transverse momentum in the image unchanged.
\end{enumerate}

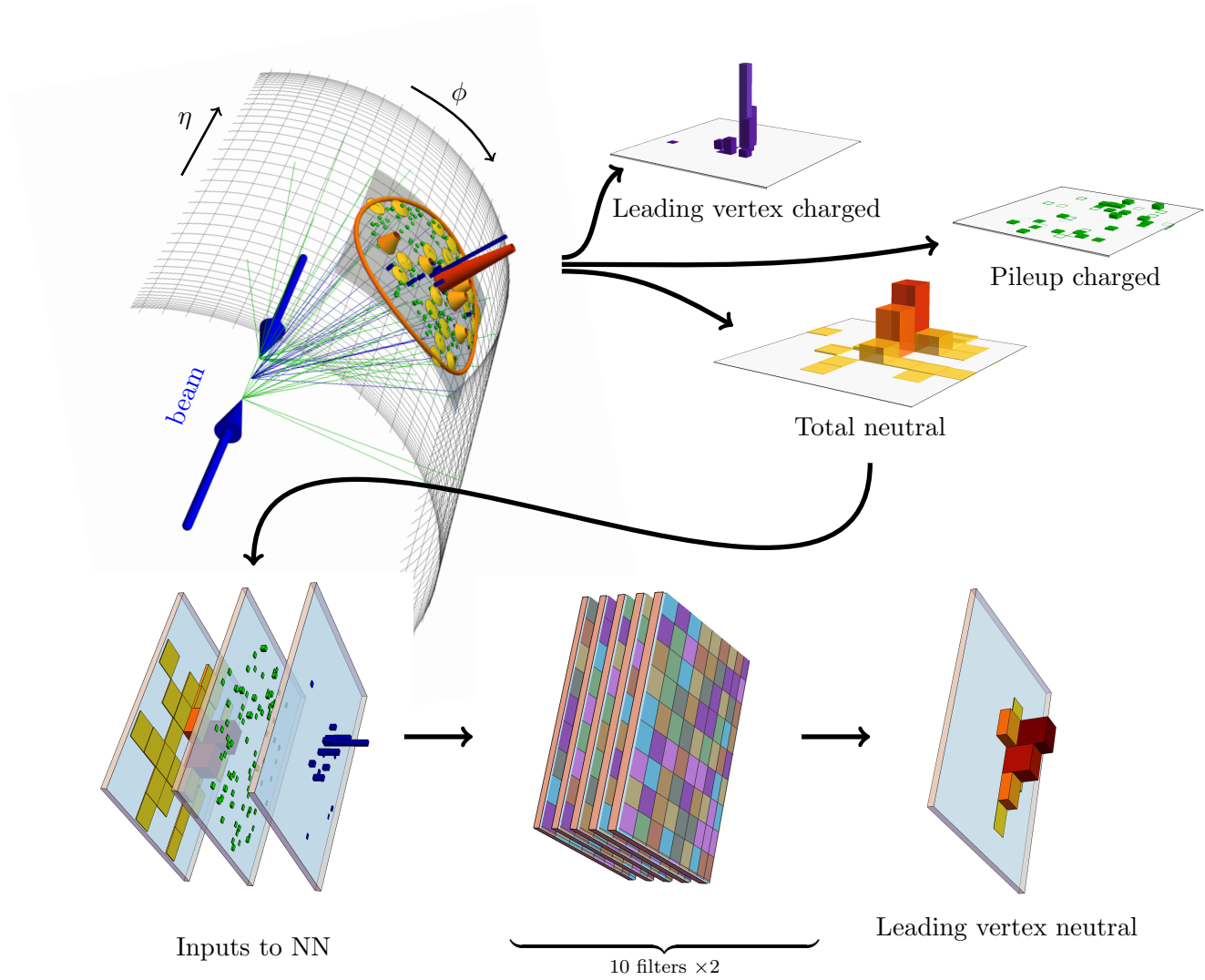
\begin{figure}[t]
\vspace{15mm}
\hspace{-2cm}
{{
\begin{tikzpicture}
\node at (-3.8,-2.3) [rotate=10] {\includegraphics[width=0.5\columnwidth, trim = {0 0mm 0 40mm}]{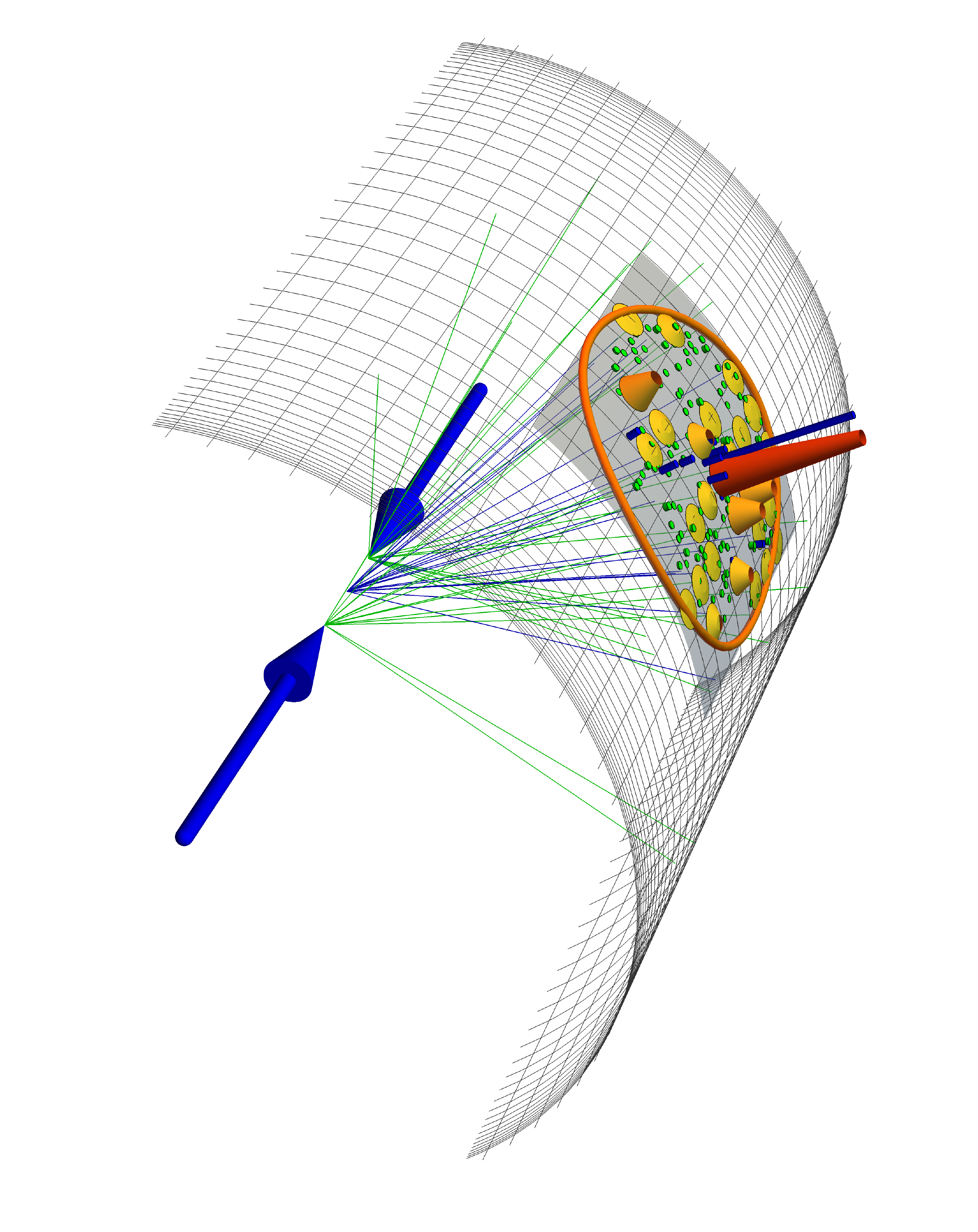}};
\node[rotate=0] at (-6,2) {$\eta$};
\draw [->,line width=1] (-6.05,1.2) to (-5.5,2.2);
\node[rotate=0] at (-2, 2.4) {$\phi$};
\draw [->,line width=1] (-2.7,2.5) to [out=-30, in = 120] (-1.5,1.4);
\node[rotate=68,,blue] at (-6,-2) {beam};
\node at (2,2) {\includegraphics[width=0.3\columnwidth]{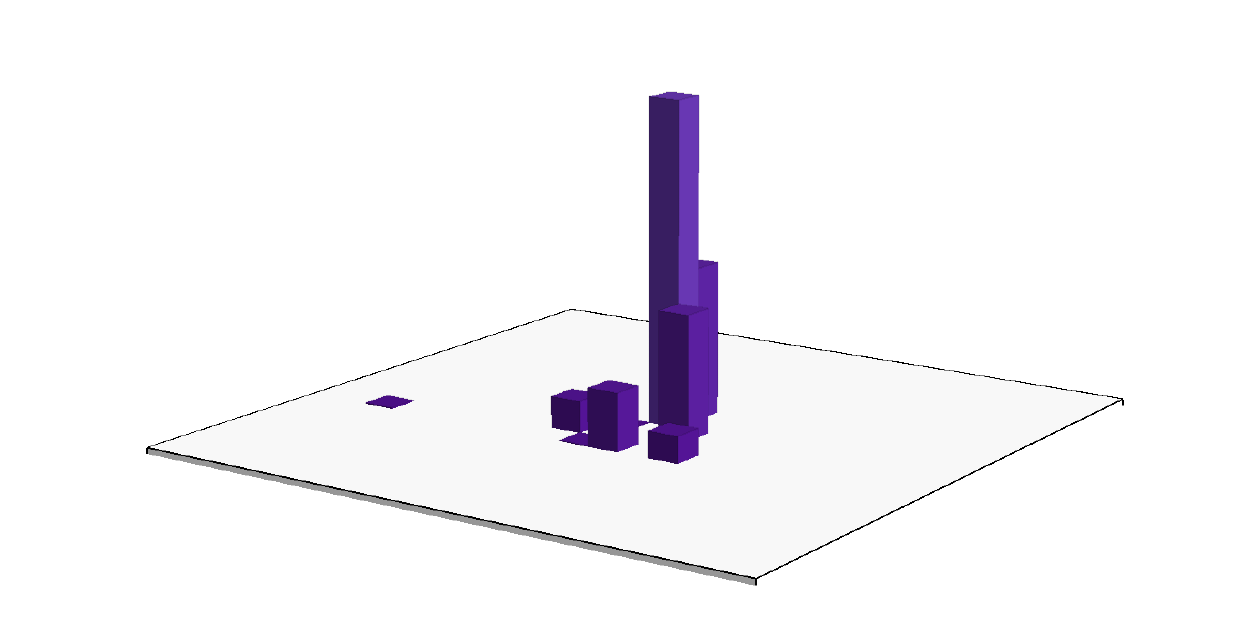}};
\node at (4,-1) {\includegraphics[width=0.3\columnwidth]{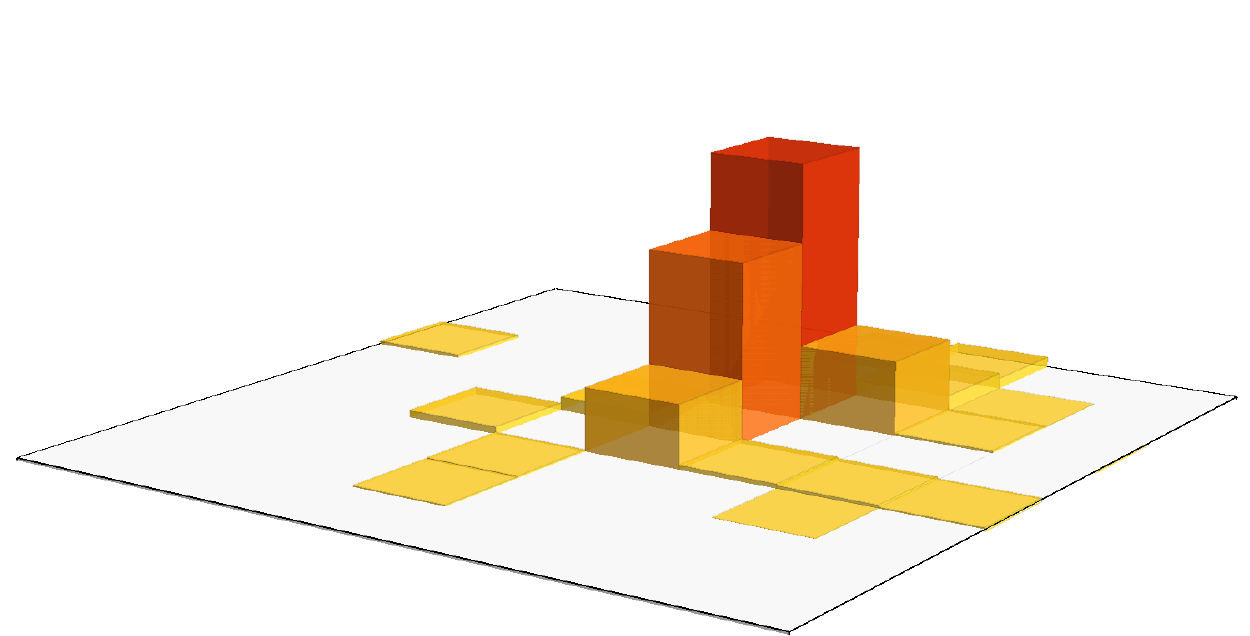}};
\node at (7,1) {\includegraphics[width=0.3\columnwidth]{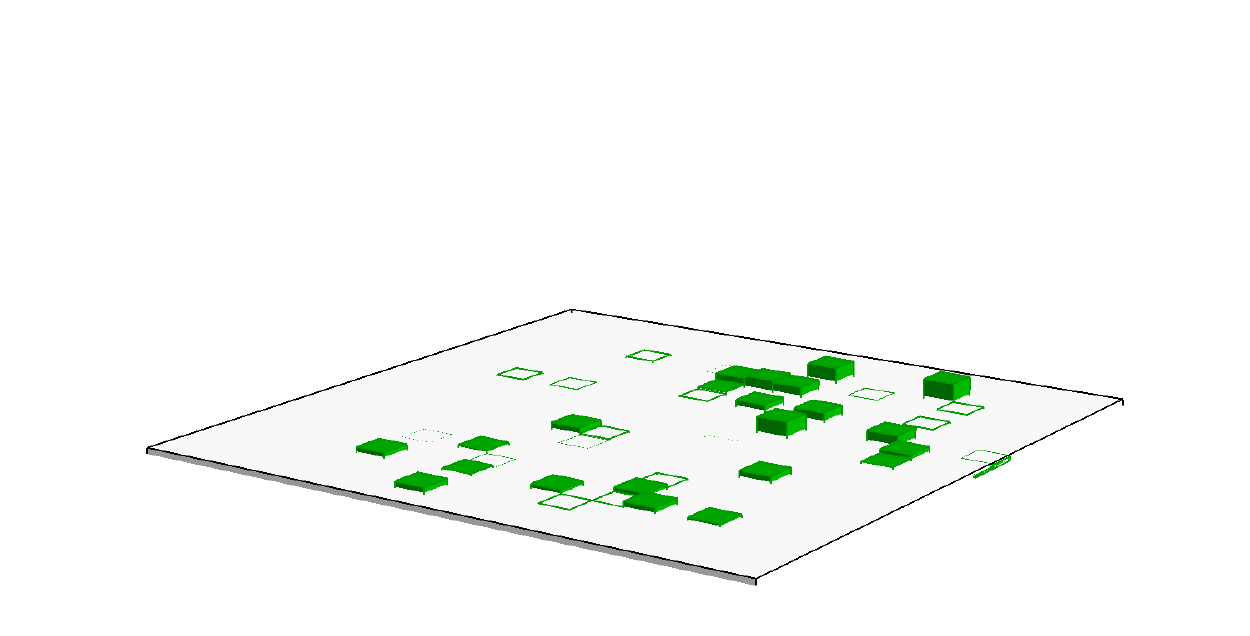}};
\draw [->,line width=2] (-0.5,0) to [out=0,in=210] (0.4,1.3);
\draw [->,line width=2] (-0.5,-0.1) to [out=0,in=190] (5,0.2);
\draw [->,line width=2] (-0.5,-0.2) to [out=0,in=150] (2,-1);
\node [below] at (2.2,1)  {Leading vertex charged};
\node [below] at (7,0)  {Pileup charged};
\node [below] at (4,-2.2)  {Total neutral};
\node at (-5,-7)  {\includegraphics[width=0.3\columnwidth]{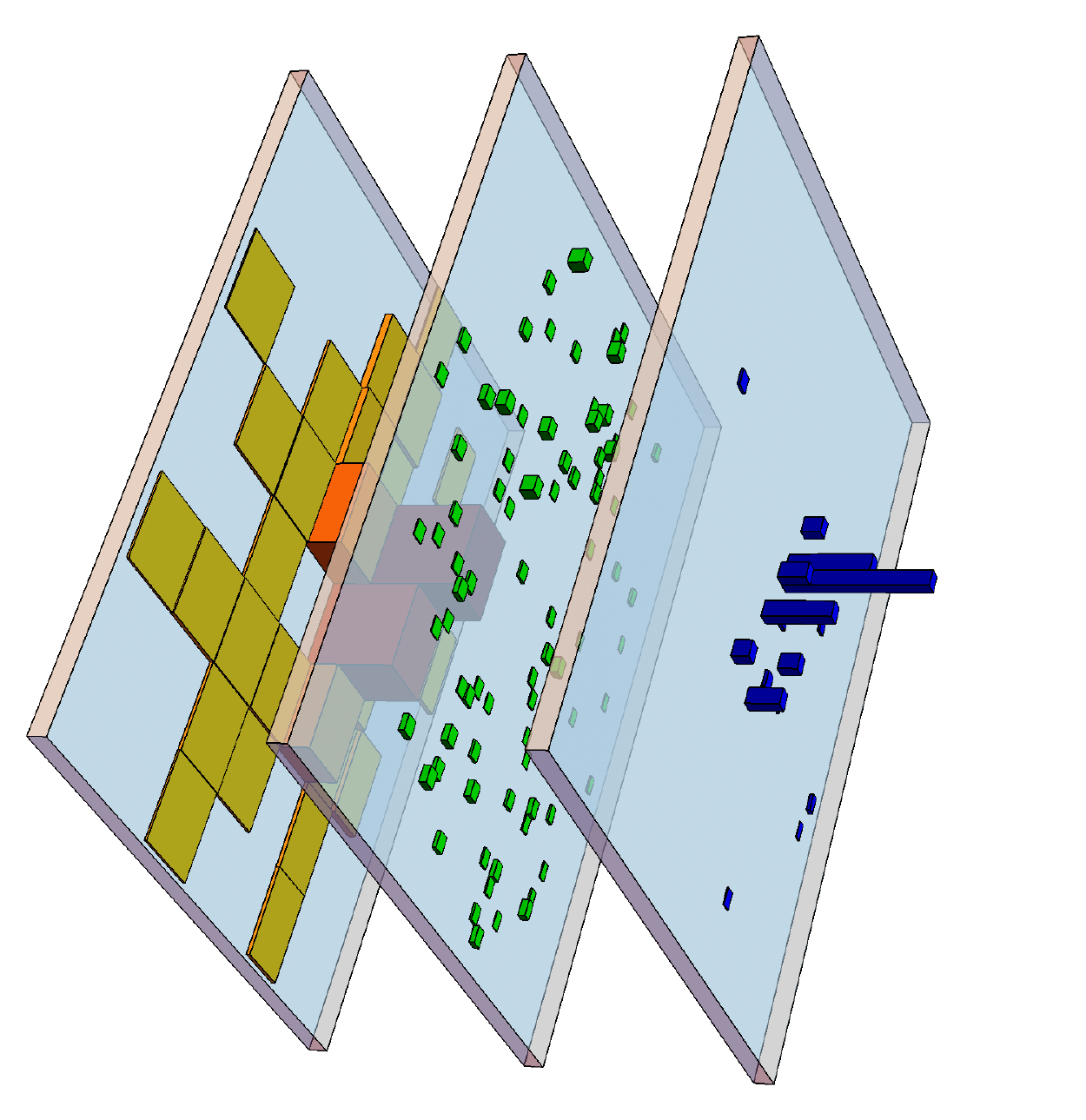}};
\node at (1,-7)  {\includegraphics[width=0.35\columnwidth,trim = {90 0 20 0}, clip]{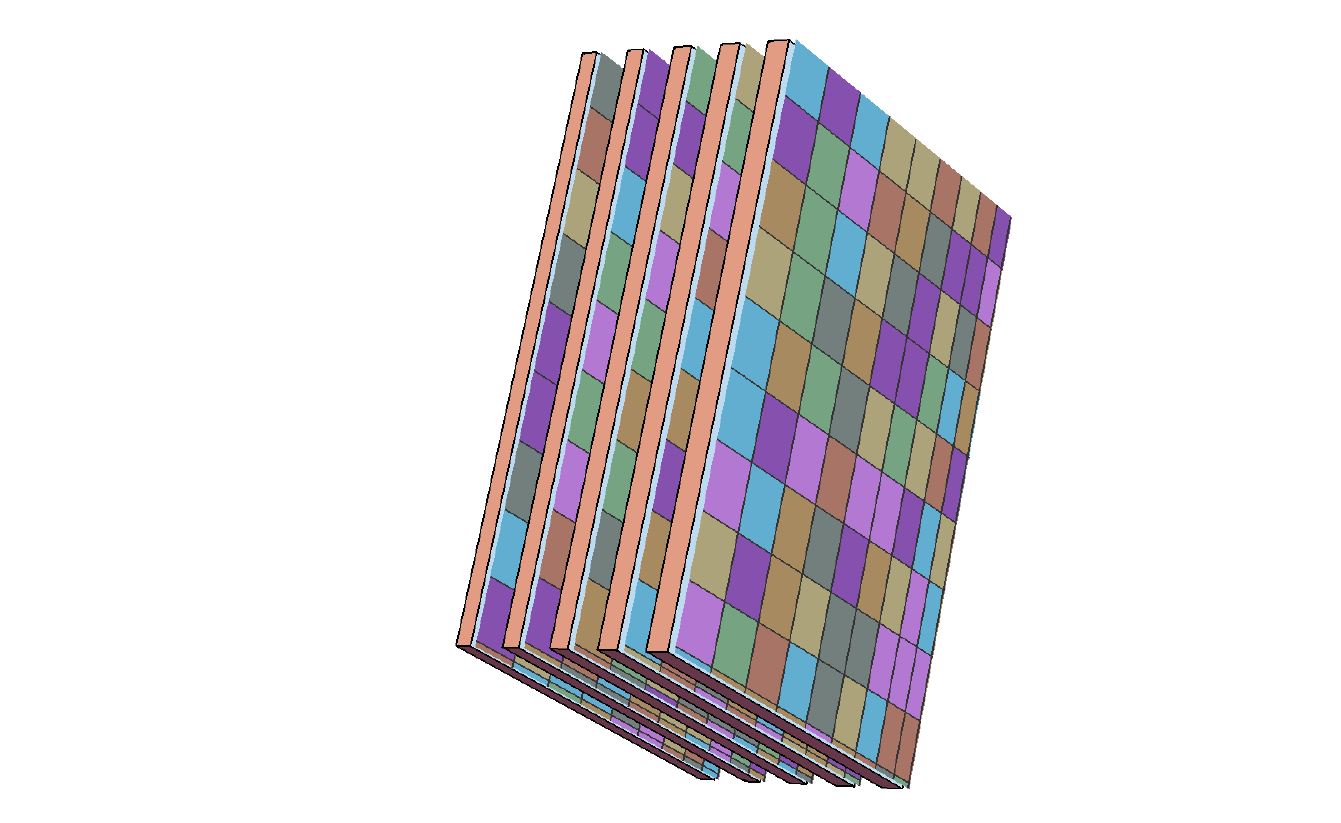}};
\node at (6,-7) {\includegraphics[width=0.15\columnwidth, trim = {5 0 5 0}, clip]{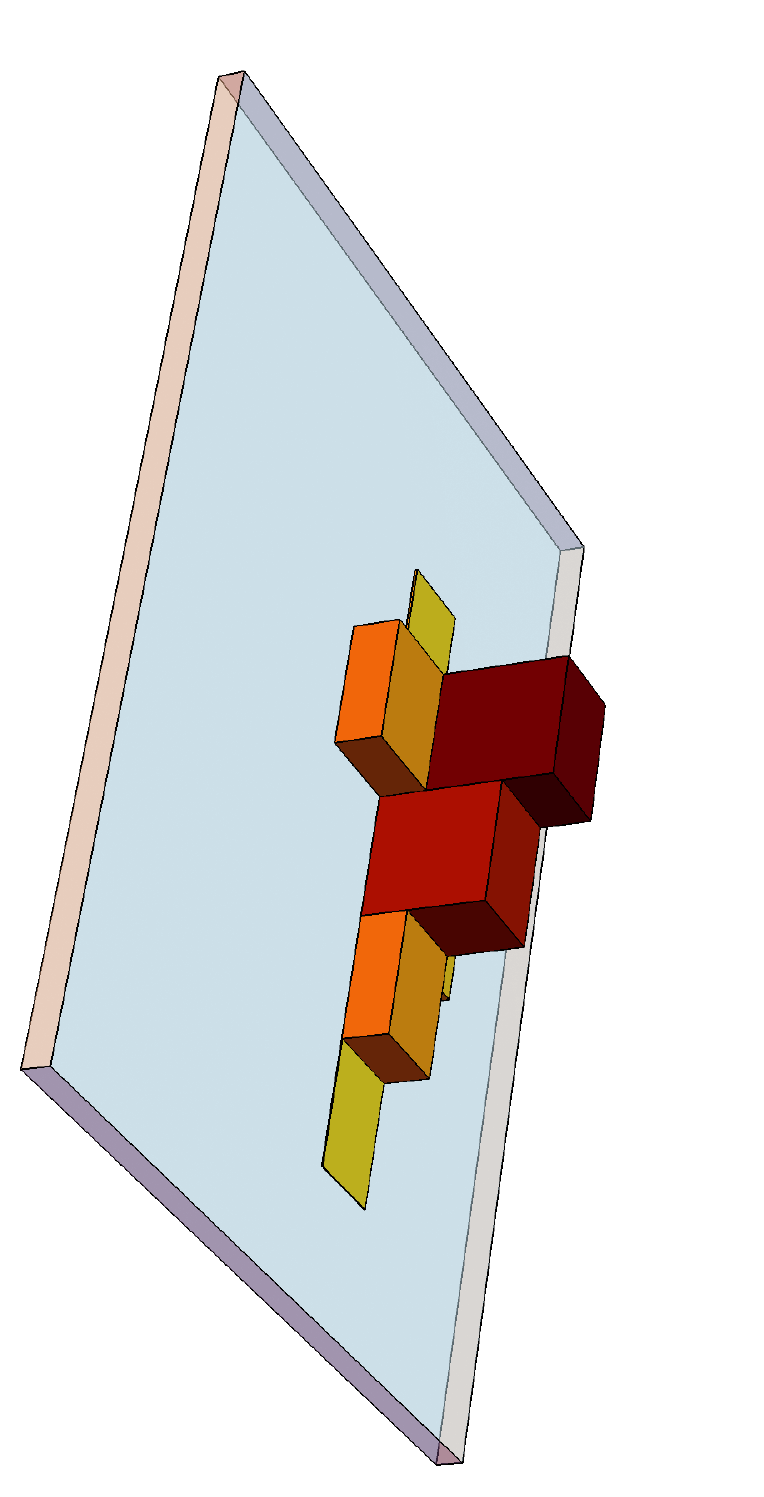}};
\node [below] at (6,-9.5)  {Leading vertex neutral};
\node [below] at (-5,-9.8)  {Inputs to NN};
\draw[->,line width=2] (4,-3) to [out=-90,in=90] (-5,-4.5);
\draw [->,line width=2] (-2.8,-7) -- (-1.8,-7);
\draw [->,line width=2] (3,-7) -- (4,-7);
\node [below] at (1,-9.8) {$\underbrace{\hspace{4.5cm}}_{\text{10 filters $\times 2$}}$};
\end{tikzpicture}
}}
\vspace{5mm}
\caption{
\label{fig:NN_arch} An illustration of the PUMML framework. The input is a three-channel image: blue/purple represents charged radiation from the leading vertex, green is charged pileup radiation, and yellow/orange/red is the total neutral radiation. The resolution of the charged images is higher than for the neutral one. These images are fed into a convolutional layer with several filters whose value at each pixel is a function of a patch around that pixel location in the input images. The output is an image combining the pixels of each filter to one output pixel.}
\end{figure}

The convolutional neural net architecture used in this study took as input $36\times 36$ pixel, three-channel pileup images. Two convolutional layers, each with 10 filters of size $6\times 6$ with $2\times2$ strides, were used after zero-padding the input images and first convolutional layer with a 2-pixel buffer on all sides. The output of the second layer has size $9\times9\times10$, with the $9\times9$ part corresponding to the size of the target output and the 10 corresponding to the number of filters in the second layer. In order to project down to a $9\times9\times1$ output, a third convolution layer with filter size $1\times1$ is used.  This last $1\times 1$ convolutional layer is a standard scheme for dimensionality reduction.  A rectified linear unit (ReLU) activation function was applied at each stage. A schematic of the framework and architecture is shown in Fig.~\ref{fig:NN_arch}.

All neural network implementation and training was performed with the python deep learning libraries Keras~\cite{keras} and Theano~\cite{theano}. The dataset consisted of 56k pileup images, with a 90\%/10\% train/test split. He-uniform initialization~\cite{heuniform} was used to initialize the model weights. The neural network was trained using the Adam~\cite{adam} algorithm with a batch size of 50 over 25 epochs with a learning rate of 0.001. The choice of loss function implicitly determines a preference for accuracy on harder pixels or softer pixels. To that end, the loss function used to train PUMML was a modified per-pixel logarithmic squared loss:
\begin{equation}\label{eq:loss}
\ell = \left< \log\left(\frac{p_T^{\text{(pred)}} + \bar p}{p_T^{\text{(true)}} + \bar p}\right)^2 \right>,
\end{equation}
where $\bar p$ is a hyperparameter that controls the choice between favoring all $p_T$ equally $(\bar p\to \infty)$ or favoring soft pixels $(\bar p \to 0)$. After mild optimization, a value of $\bar p = 10$ GeV was chosen, though the performance of the model as measured by correlations between reconstructed and true observables is relatively robust to this choice. PUMML was found to give good performance even with a standard loss function such as the mean squared error, which favors all $p_T$ equally.

The PUMML architecture is \emph{local} in that the rescaling of a neutral pixel is a function solely of the information in a patch in $(\eta,\phi)$-space around that pixel. The size of this patch can be controlled by tuning the filter sizes and number of layers in the architecture. Further, due to weight-sharing in convolutional layers, the same function is applied for all pixels. Building this locality and translation invariance into the architecture ensures that the algorithm learns a universal pileup mitigation technique, while carrying the benefit of drastically reducing the number of model parameters. Indeed, the PUMML architecture used in this study has only 4,711 parameters, which is small on the scale of deep learning architectures, but serves to highlight the effectiveness of using modern machine learning techniques (such as convolutional layers) in high energy physics without necessarily using large or deep networks.

While we considered jets and jet images in this study, the PUMML architecture using convolutional nets readily generalizes to event-level applications. The locality of the algorithm implies that the trained model can be applied to any desired region of the event using only the surrounding pixels. To train the model on the event level, either the existing PUMML architecture could be generalized to larger inputs and outputs or the event could be sliced into smaller images and the model trained as in the present study. The parameters of the PUMML architecture are the convolutional filter sizes, the number of filters per layer, and the number of convolutional layers, which may be optimized for a specific application. Here, we have presented an architecture optimized for simplicity and performance for jet-level pileup subtraction. PUMML is designed to be applicable at both jet- and event-level.

\section{Performance\label{sec:performance}}

To test the PUMML algorithm, we consider $q\bar q$ light-quark-initiated jets coming from the decay of a scalar with mass $m_\phi = 500$ GeV. Events were generated using Pythia 8.183~\cite{Pythia8.2:2015} with the default tune for $pp$ collisions at $\sqrt{s} = 13$ TeV.  Pileup was generated by overlaying soft QCD processes onto each event. Final state particles except muons and neutrinos were kept. The events were clustered with FastJet 3.1.3~\cite{FastJet:2012} using the anti-$k_t$ algorithm~\cite{cacc2008} with a jet radius of $R = 0.4$. A parton-level $p_T$ cut of 95 GeV was applied and up to two leading jets with $p_T > 100$ GeV and $\eta\in[-2.5,2.5]$ were selected from each event. All particles were taken to be massless.

\begin{figure}[t]
\centering
\includegraphics[scale = 0.55]{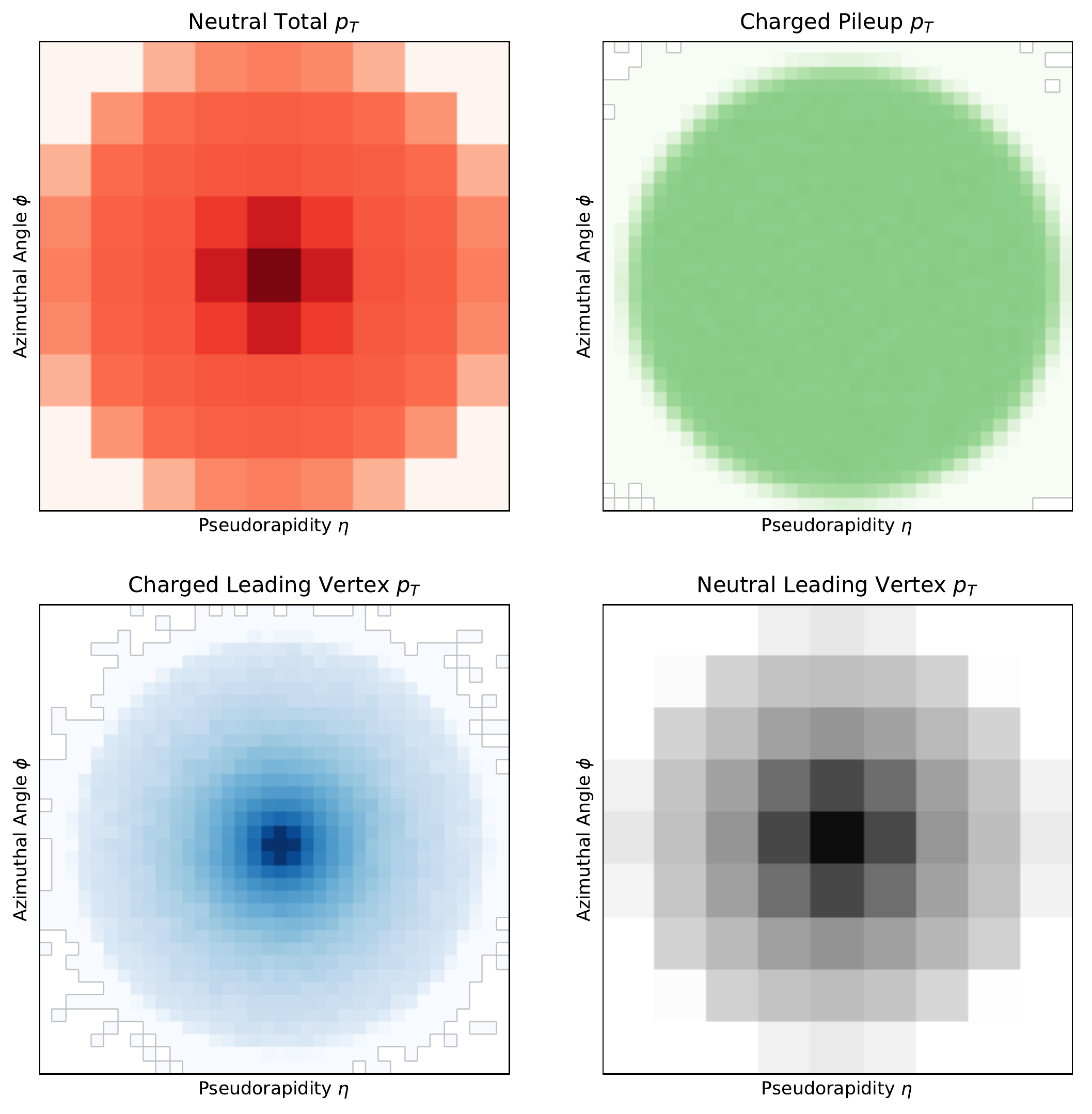}
\caption{\label{fig:jetimagesRGB} The average leading-jet images for a 500 GeV scalar decaying to light-quark jets with $\langle \text{NPU} \rangle = 140$ pileup, separated by all neutral particles (top left), charged pileup particles (top right), charged leading vertex particles (bottom left), and neutral leading vertex particles (bottom right). Different pixelizations are used for charged and neutral images to reflect the differences in calorimeter resolution. The charged and total neutral images comprise the three-channel input to the neural network, which is trained to predict the neutral leading vertex image.}
\end{figure}

\begin{figure}[t]
\centering
\begin{tikzpicture}
\node at (-6,0) {\includegraphics[width=0.23\columnwidth,trim = {0 0 20 0}, clip]{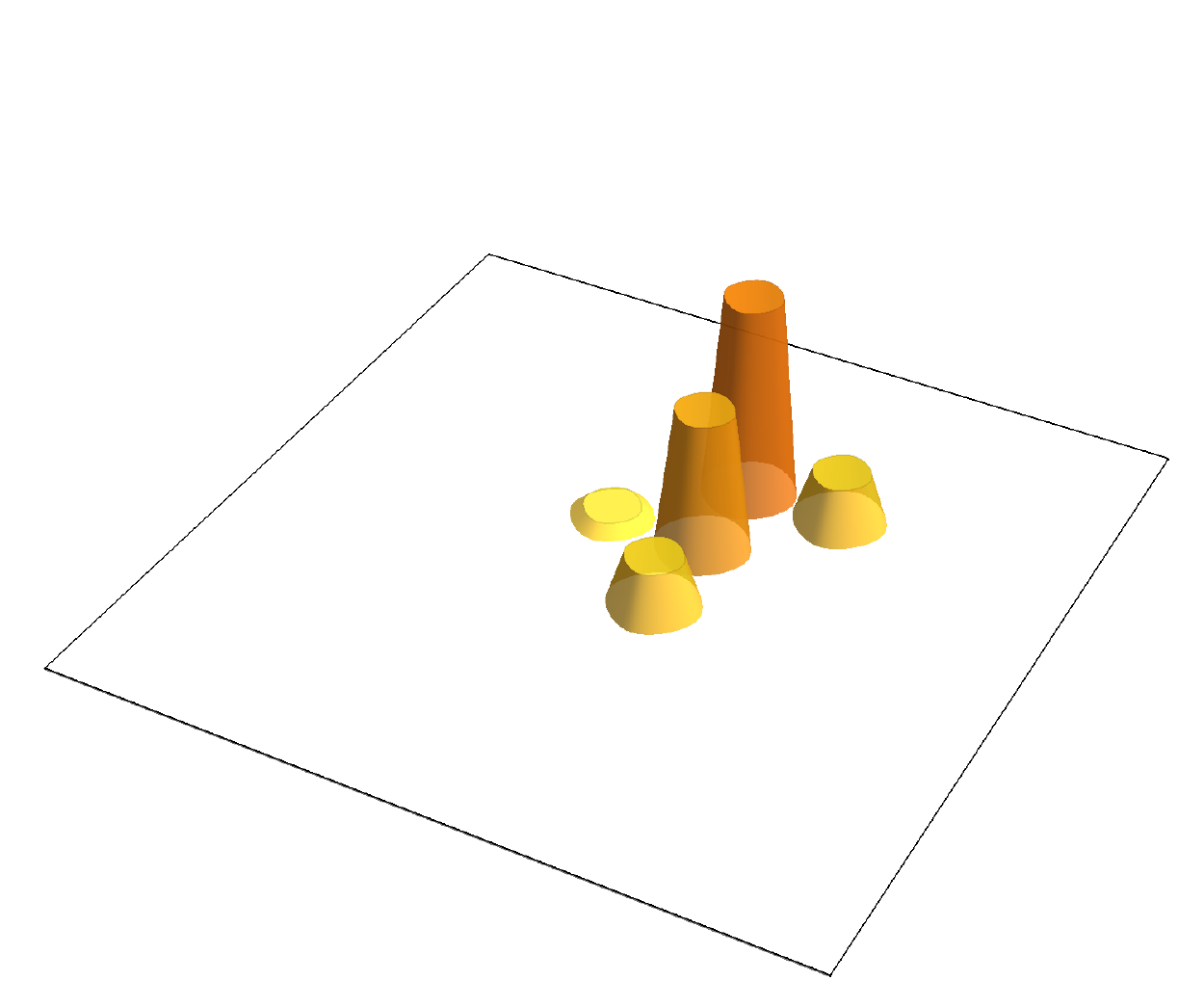}};
\node at (-3,0) {\includegraphics[width=0.22\columnwidth,trim = {40 0 20 0}, clip]{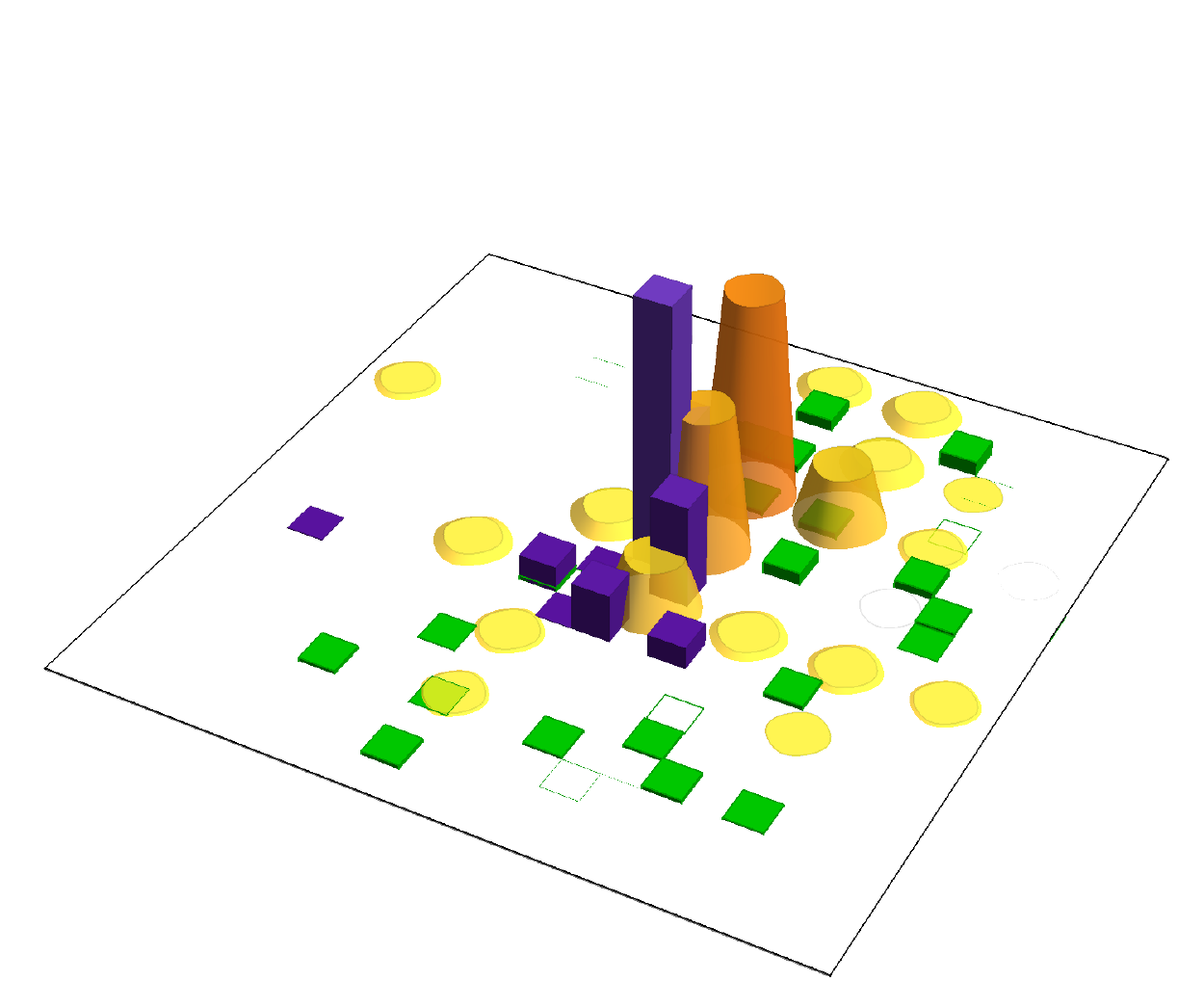}};
\node at (0,0) {\includegraphics[width=0.2\columnwidth,trim = {60 0 20 0}, clip]{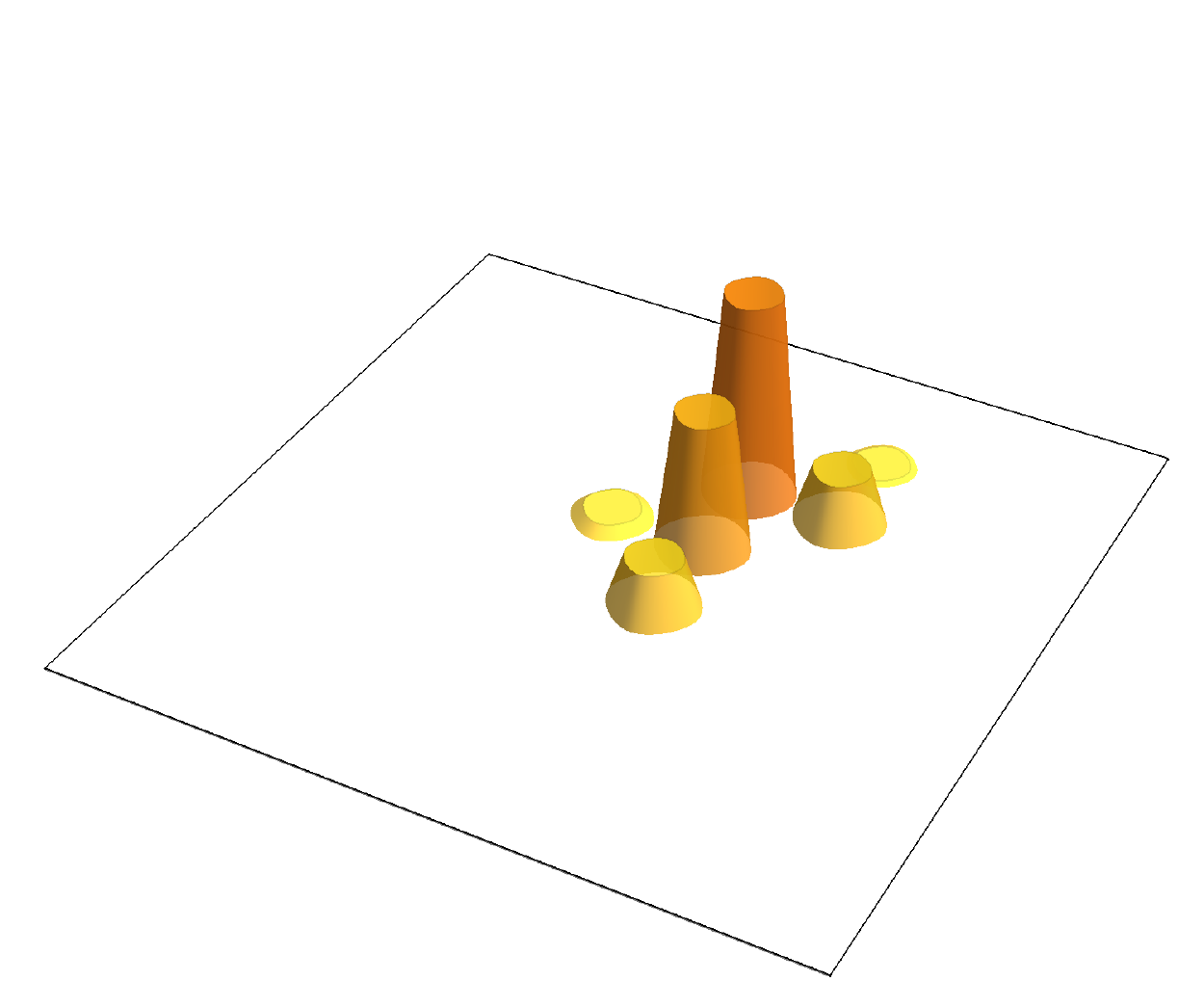}};
\node at (3,0) {\includegraphics[width=0.2\columnwidth,trim = {40 0 20 0}, clip]{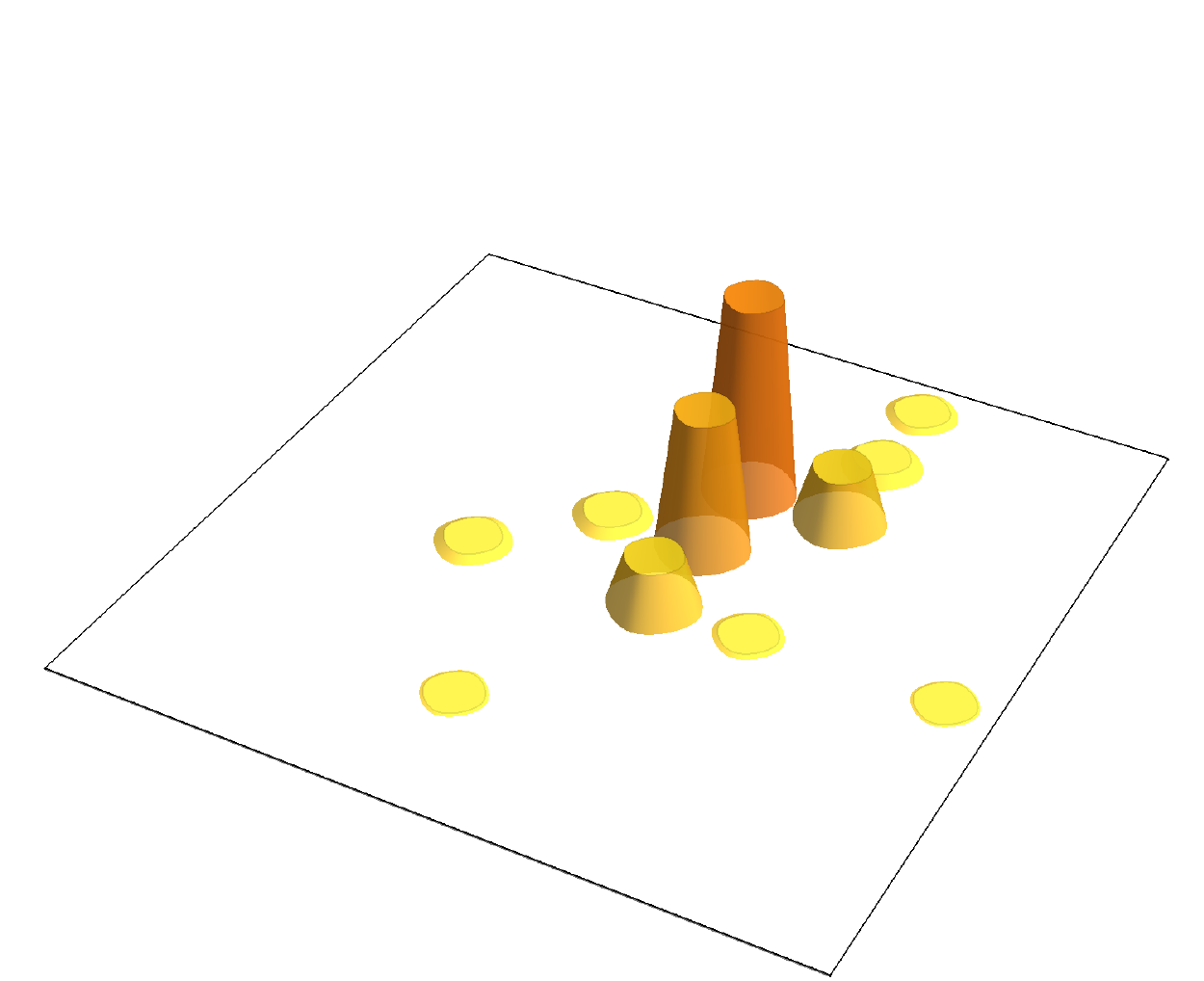}};
\node at (6,0) {\includegraphics[width=0.21\columnwidth,trim = {40 0 0 0}, clip]{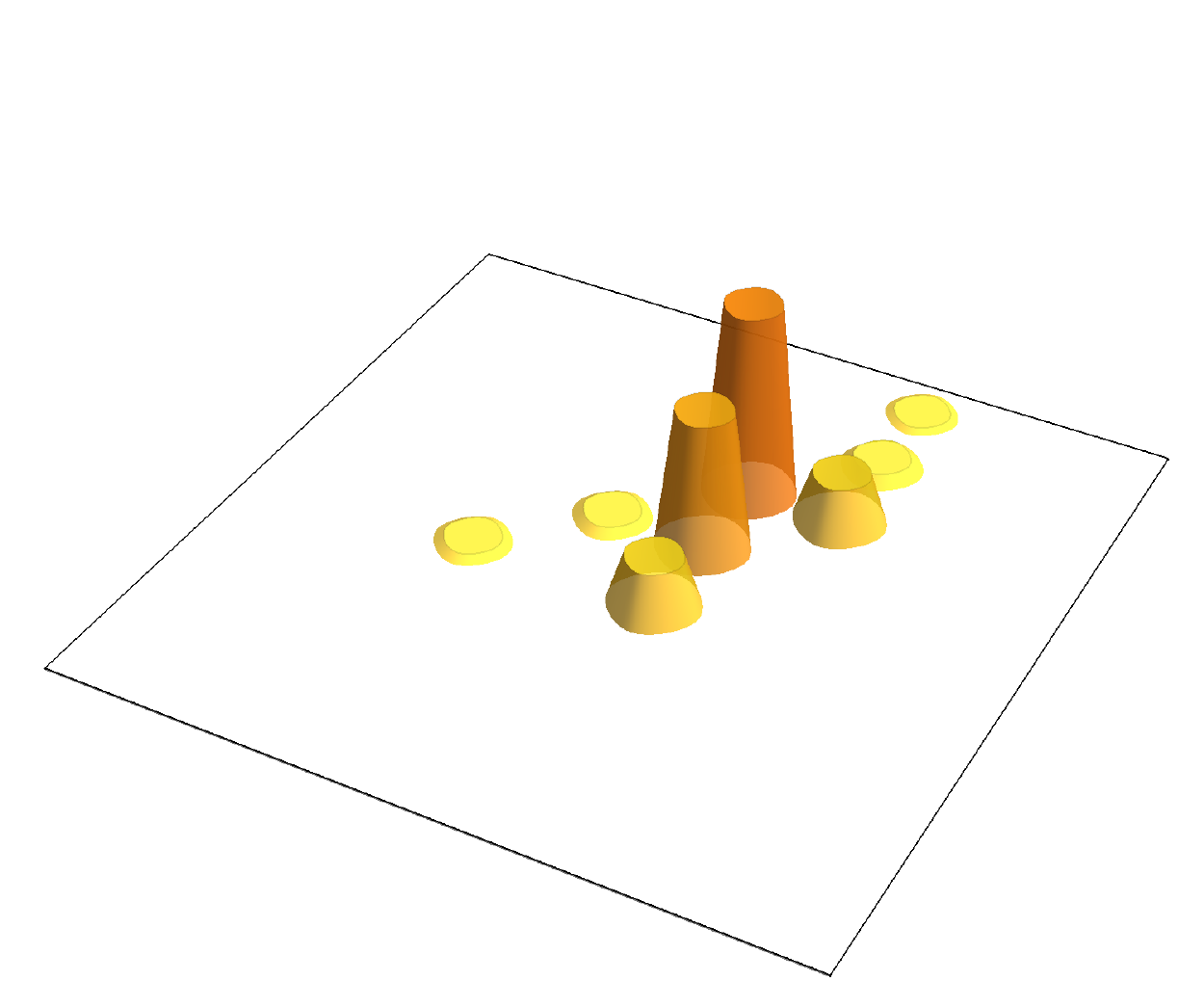}};
\node at (-6,1.5) {Leading Vertex};
\node at (-3,1.5) {with Pileup};
\node at (-0,1.5) {PUMML};
\node at (3,1.5) {PUPPI};
\node at (6,1.5) {SoftKiller};
\node at (-6,-3) {\includegraphics[width=0.23\columnwidth,trim = {0 0 20 0}, clip]{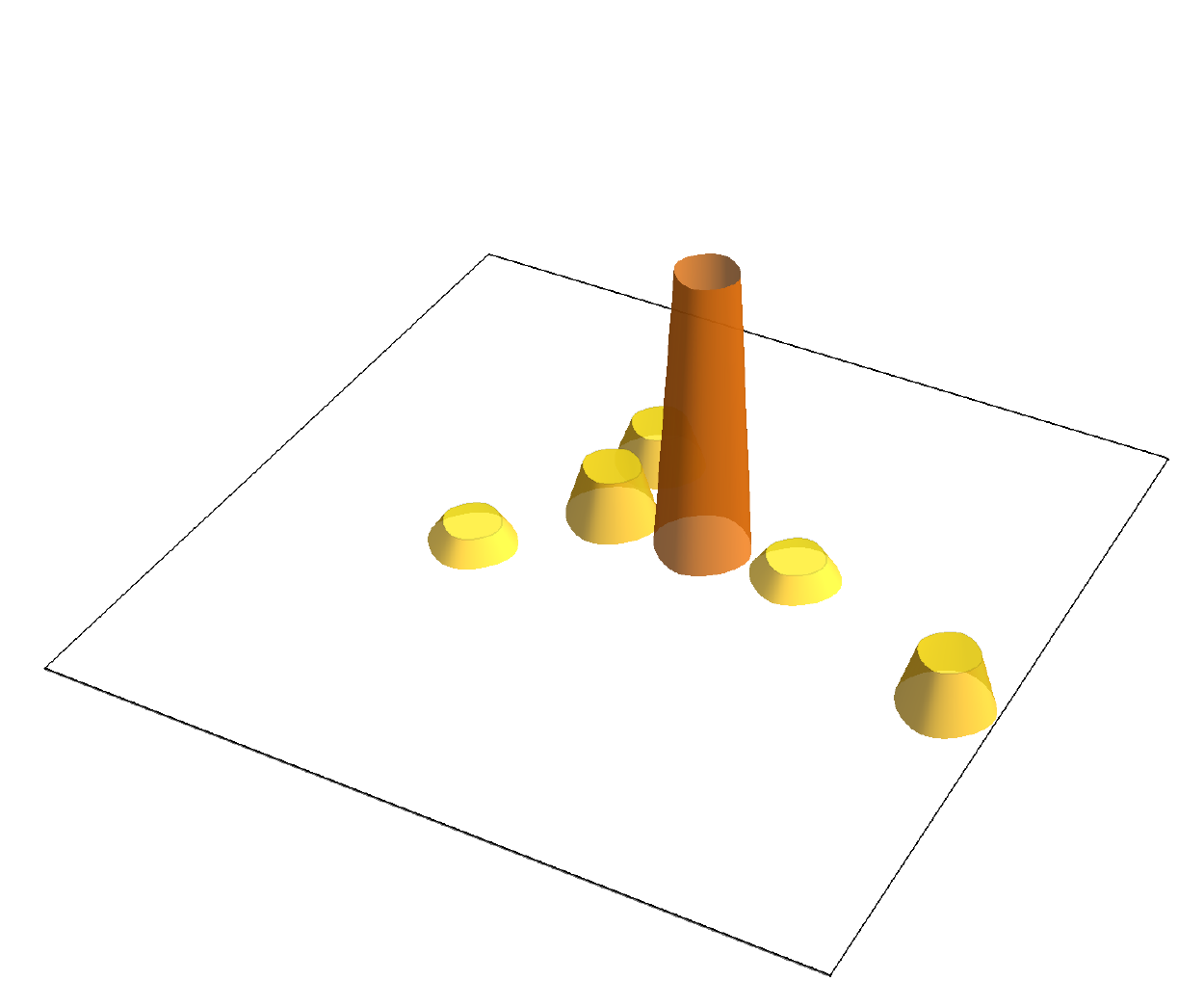}};
\node at (-3,-3) {\includegraphics[width=0.22\columnwidth,trim = {40 0 20 0}, clip]{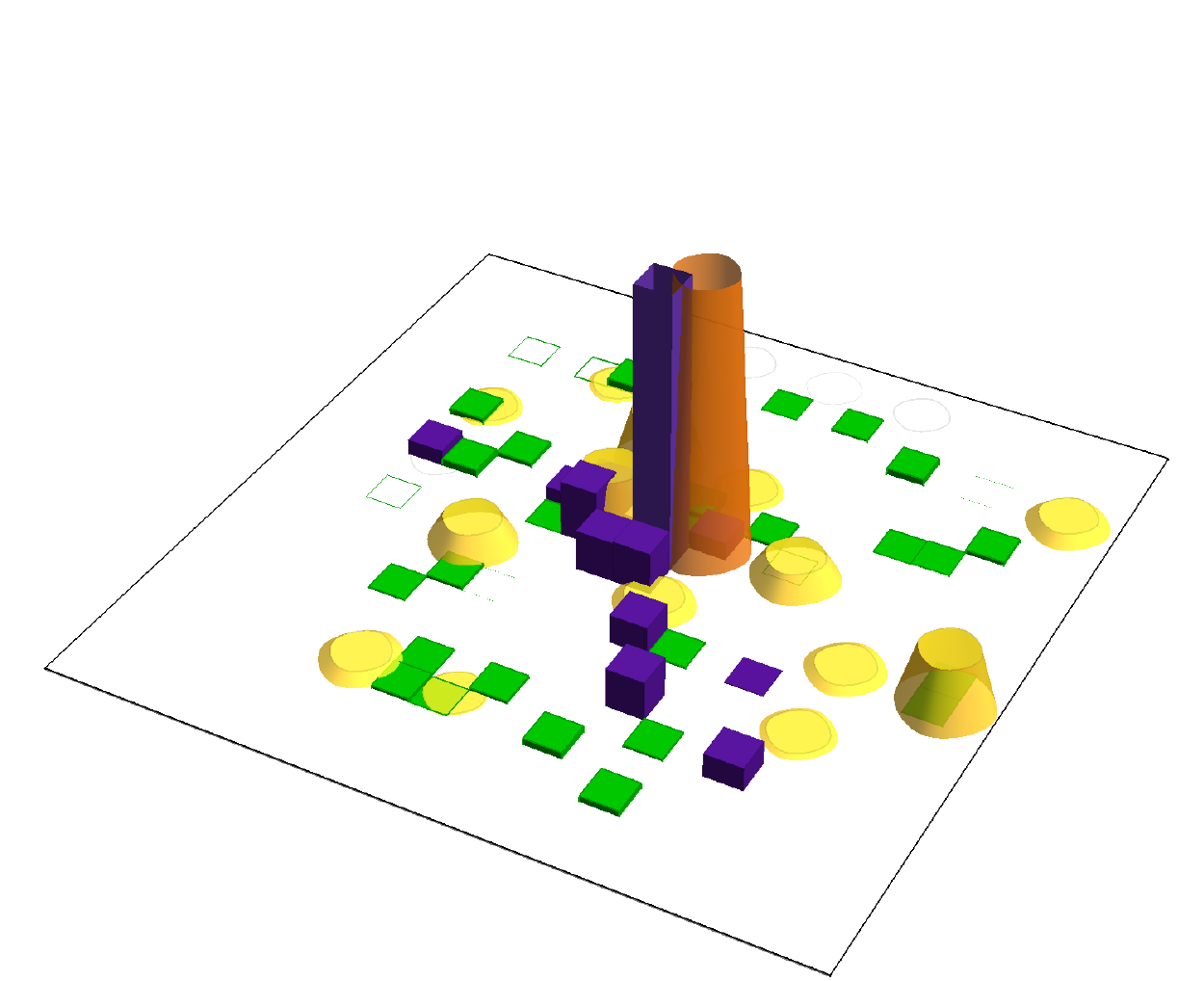}};
\node at (0,-3) {\includegraphics[width=0.2\columnwidth,trim = {60 0 20 0}, clip]{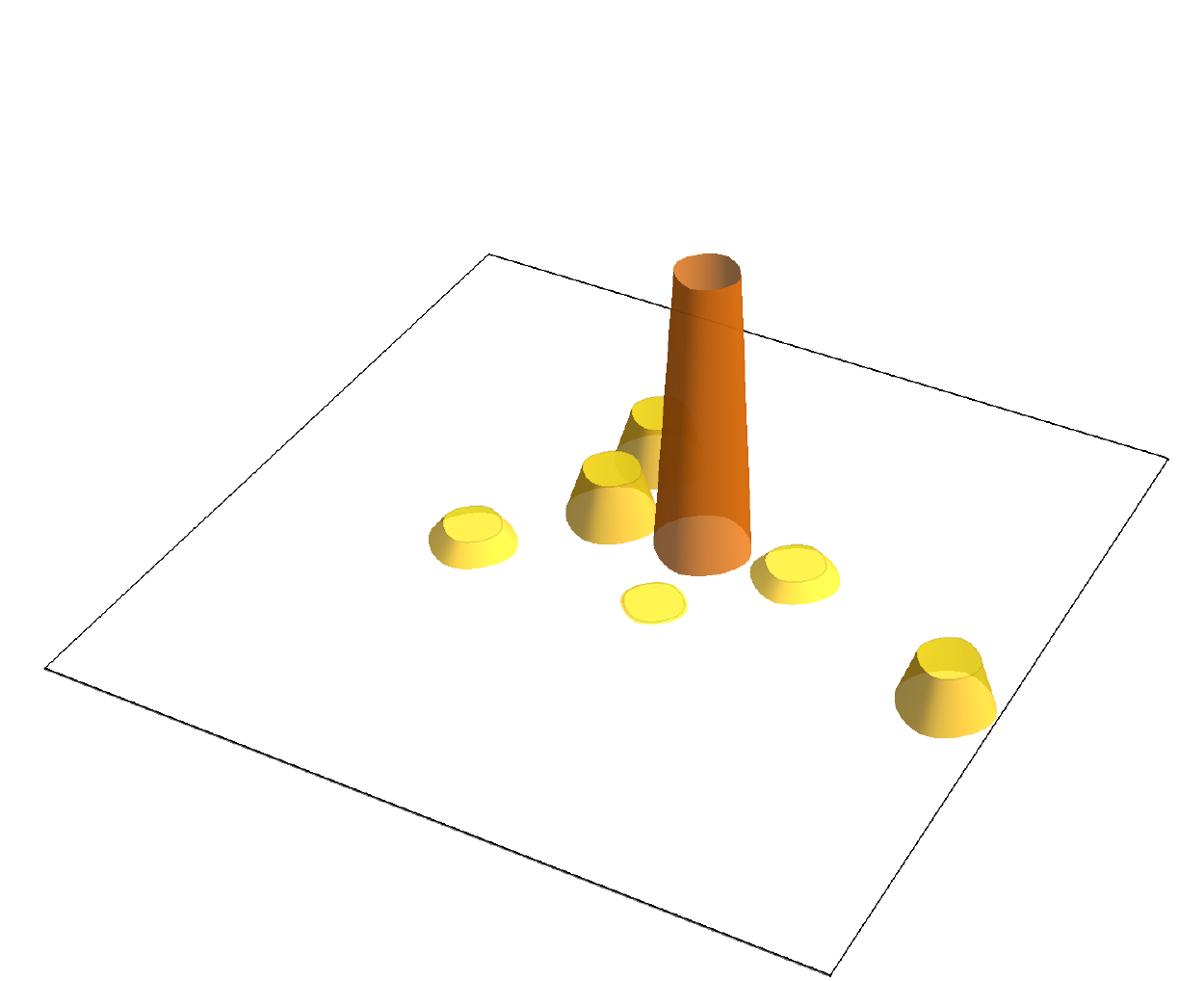}};
\node at (3,-3) {\includegraphics[width=0.2\columnwidth,trim = {40 0 20 0}, clip]{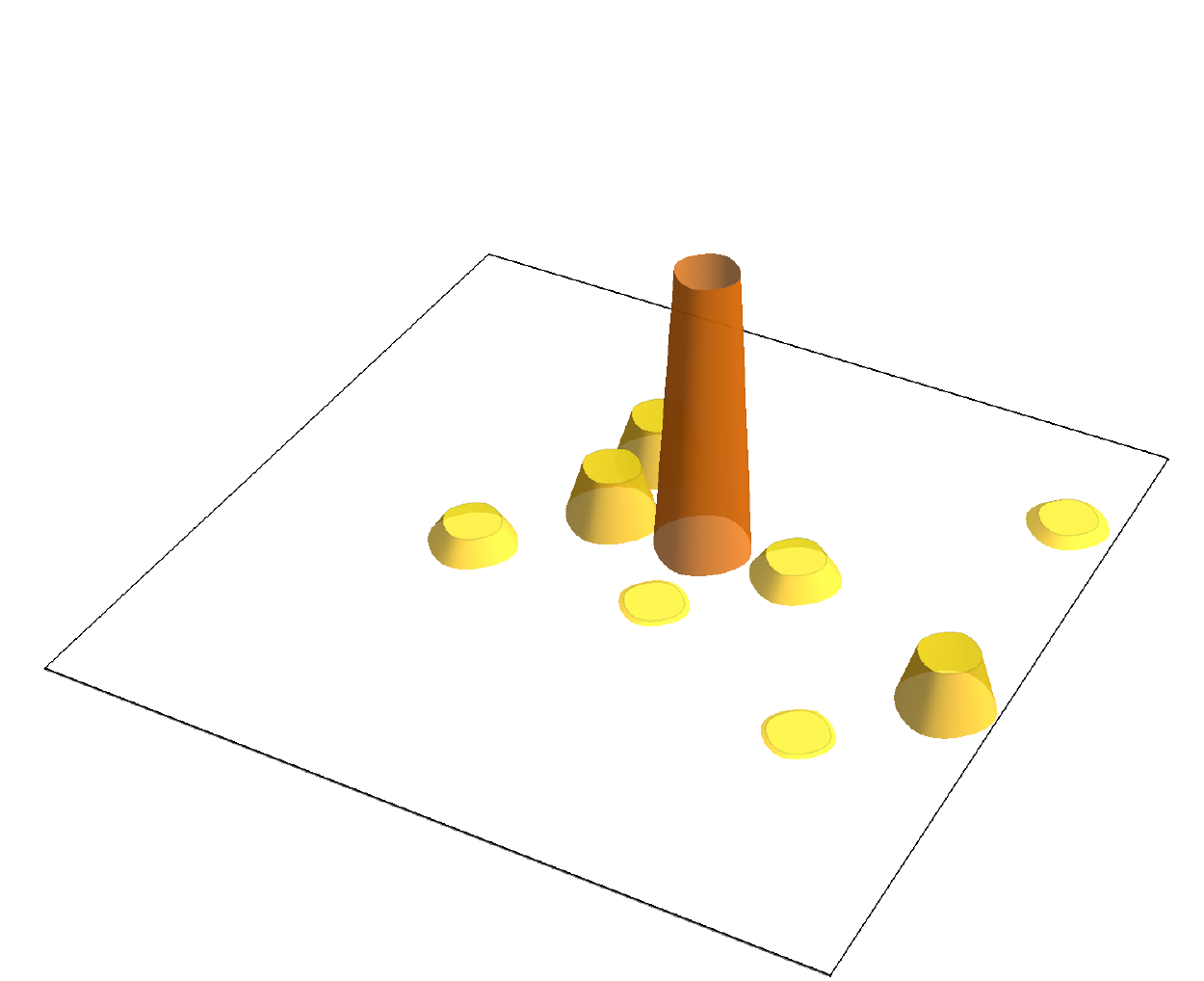}};
\node at (6,-3) {\includegraphics[width=0.21\columnwidth,trim = {40 0 0 0}, clip]{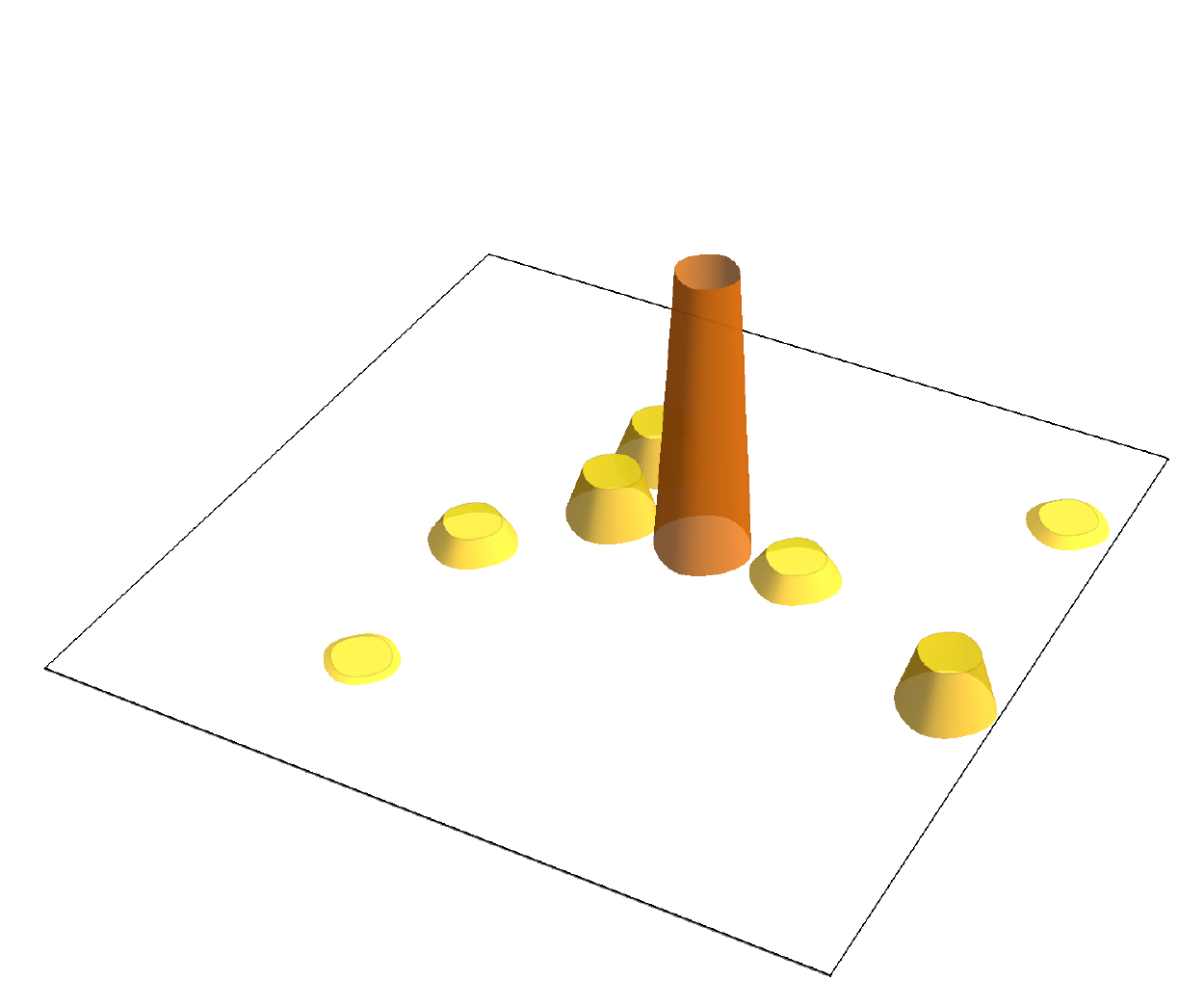}};
\node at (-6,-6) {\includegraphics[width=0.23\columnwidth,trim = {0 0 20 0}, clip]{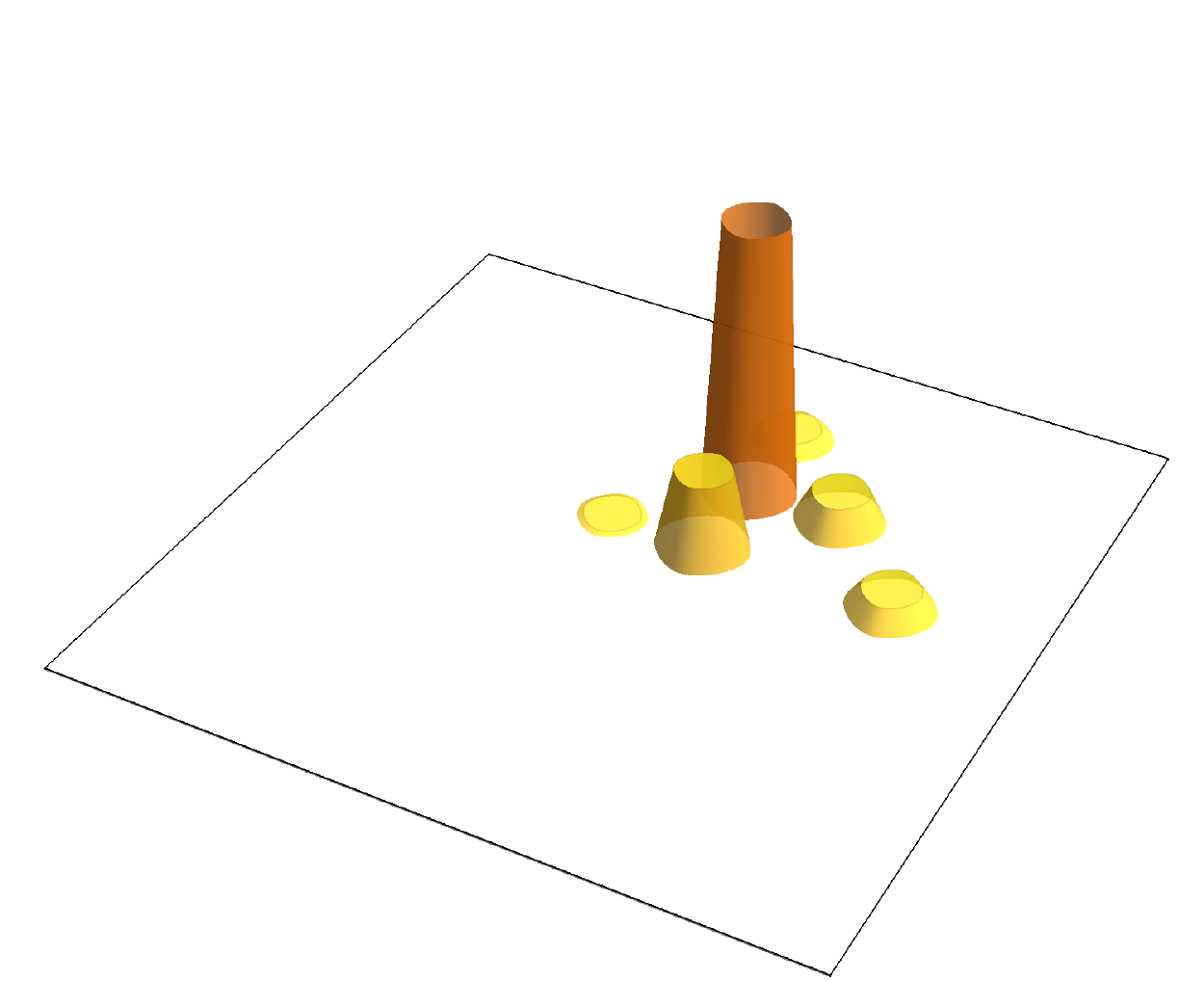}};
\node at (-3,-6) {\includegraphics[width=0.22\columnwidth,trim = {40 0 20 0}, clip]{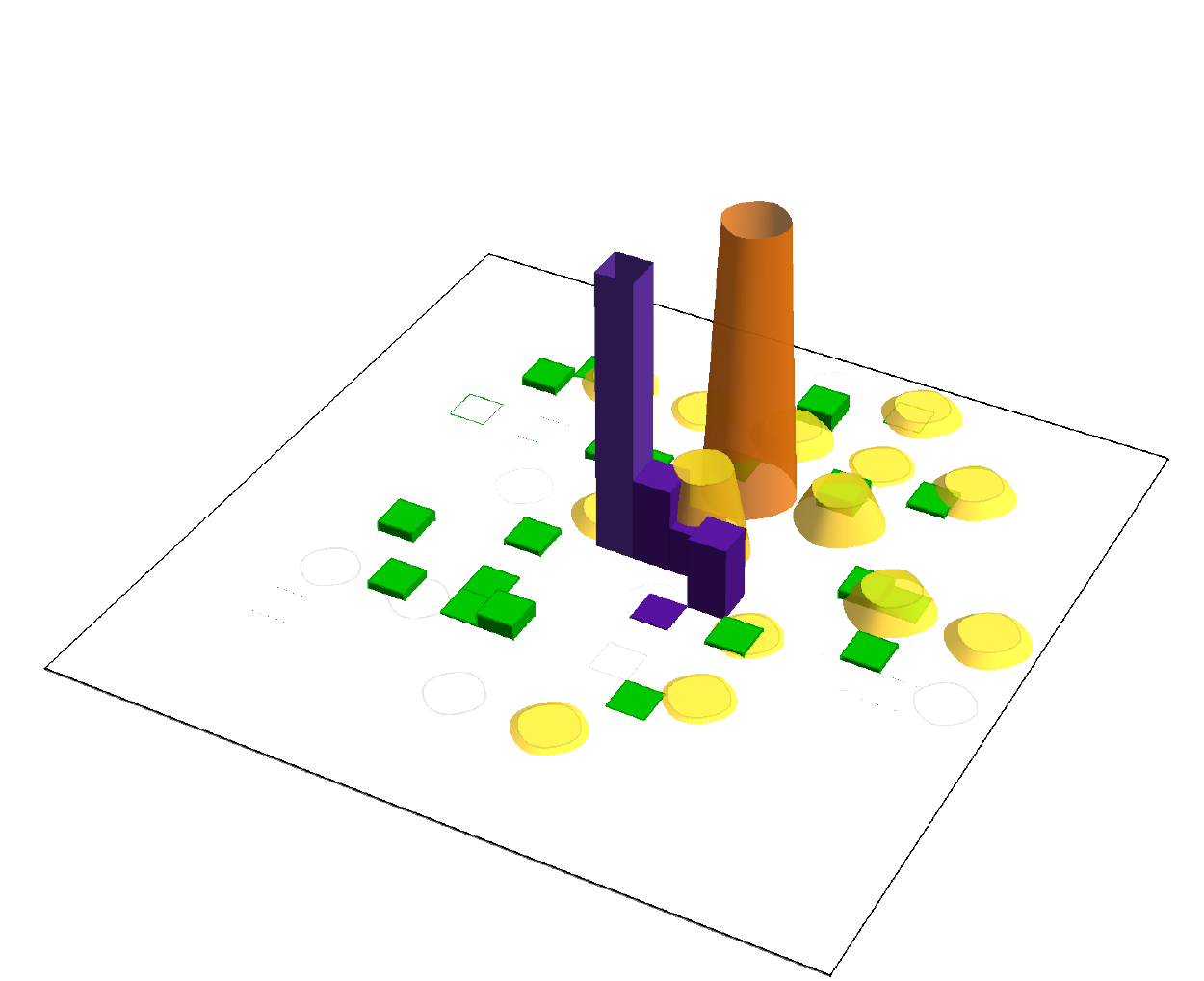}};
\node at (0,-6) {\includegraphics[width=0.2\columnwidth,trim = {60 0 20 0}, clip]{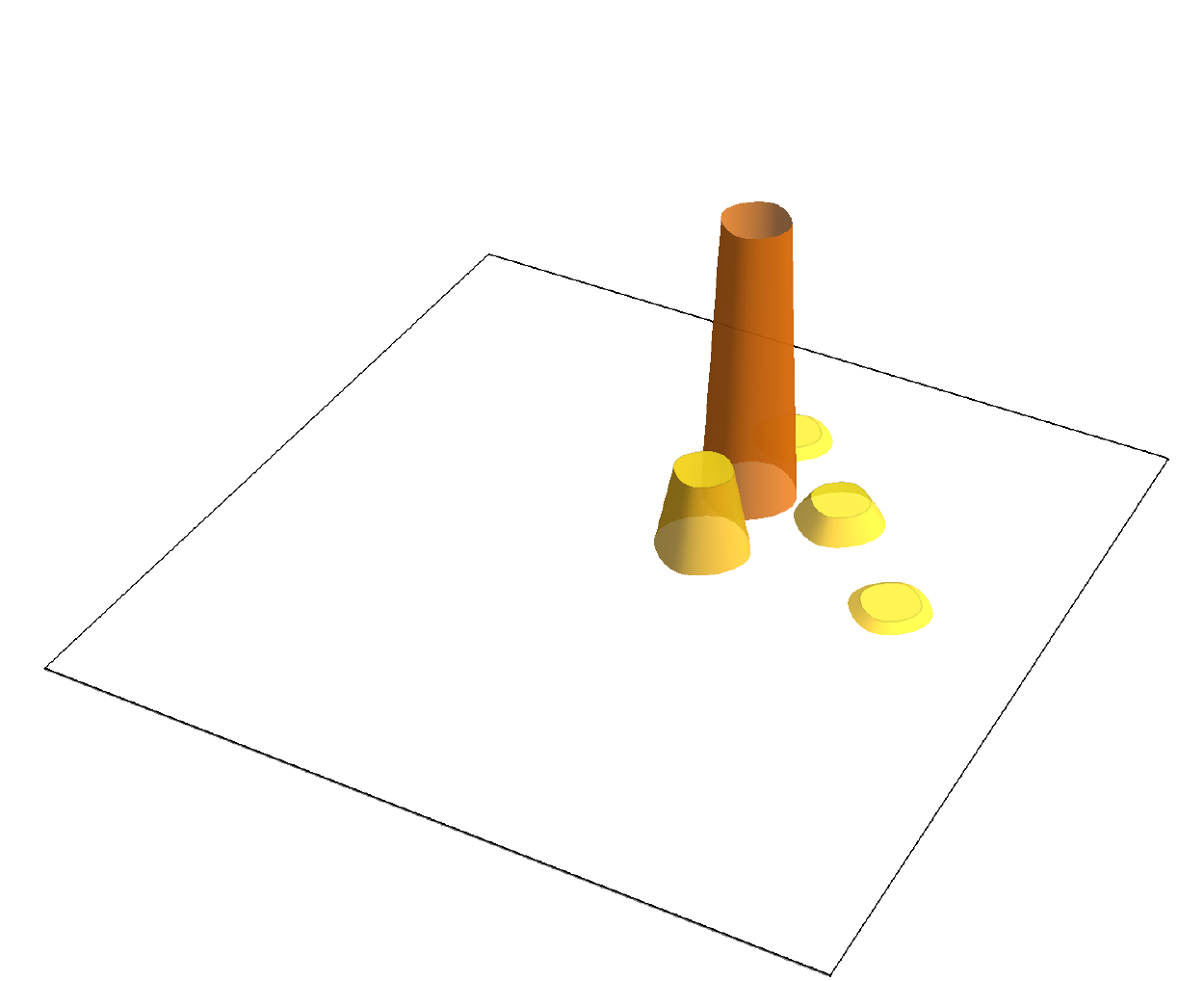}};
\node at (3,-6) {\includegraphics[width=0.2\columnwidth,trim = {40 0 20 0}, clip]{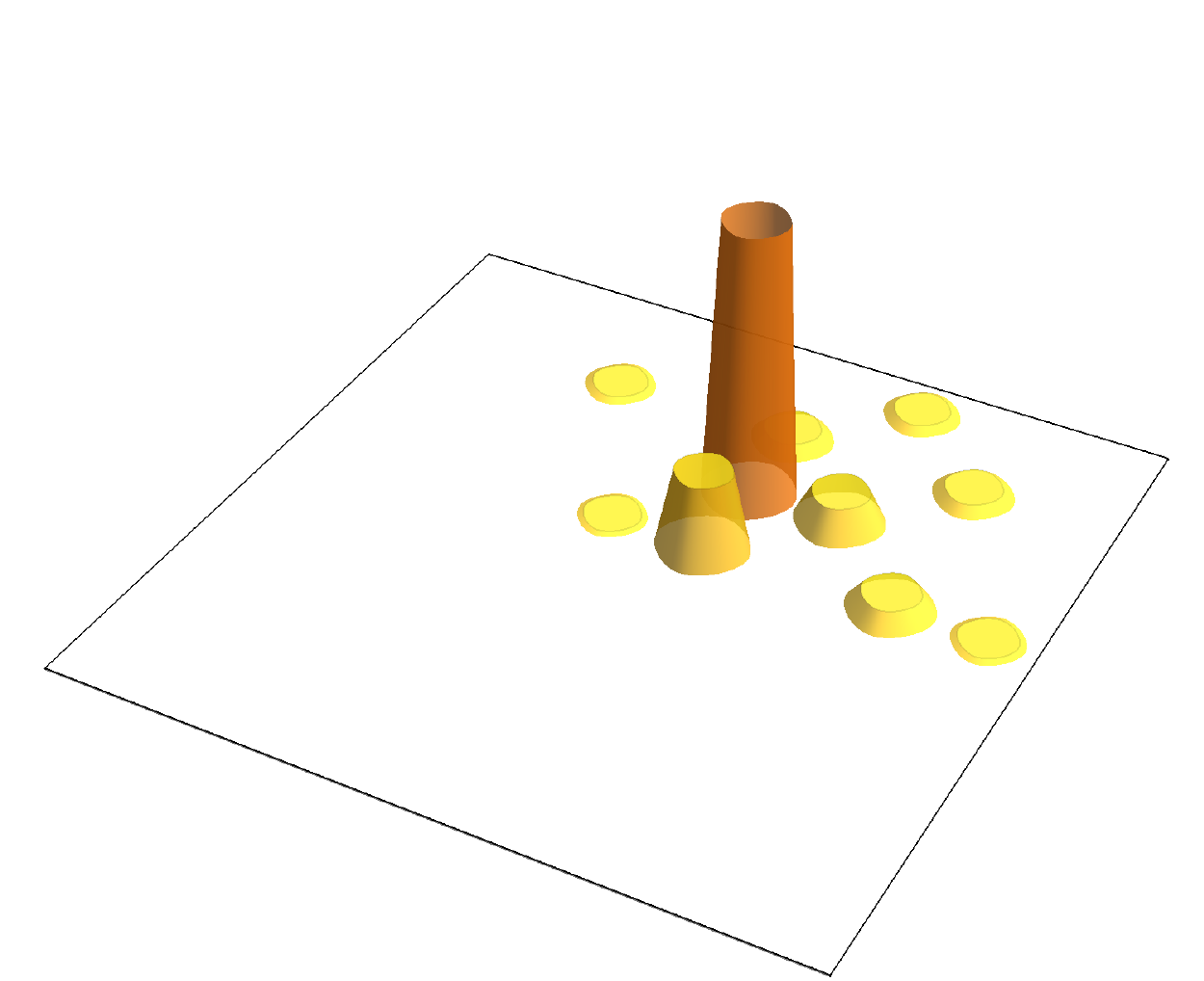}};
\node at (6,-6) {\includegraphics[width=0.21\columnwidth,trim = {40 0 0 0}, clip]{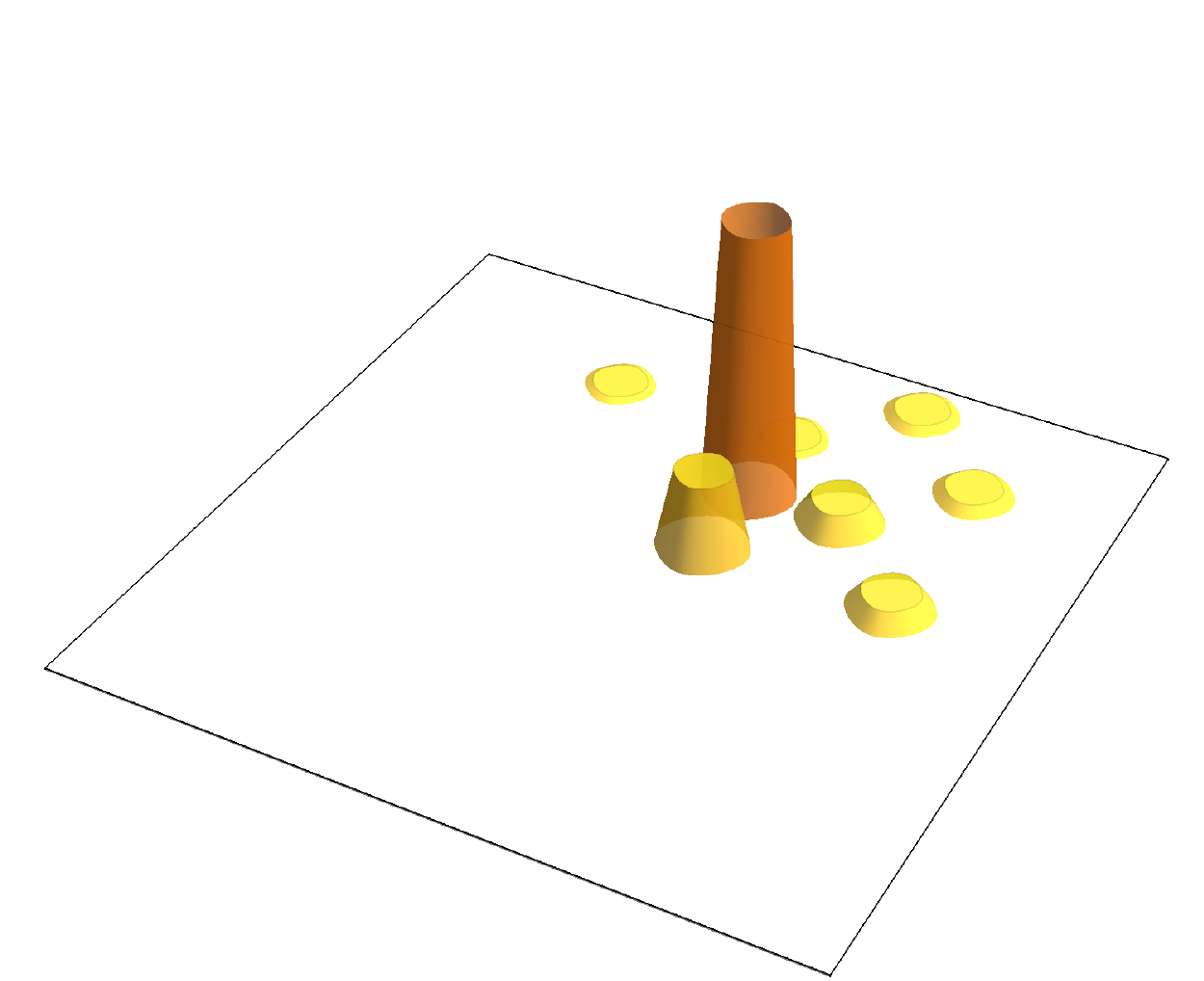}};
\end{tikzpicture}
\caption{\label{fig:ex_imgs1}  Depictions of three randomly chosen leading jets. Blue/purple represents charged radiation from the leading vertex, green is charged pileup radiation, and yellow/orange/red is the neutral radiation. Shown from left to right are the true neutral leading vertex particles, the event with pileup and charged leading vertex information, followed by the neutral leading vertex particles predicted by PUMML, PUPPI, and SoftKiller. From examining these events, it appears that PUMML has learned an effective pileup mitigation strategy.\vspace{4mm}}
\end{figure}
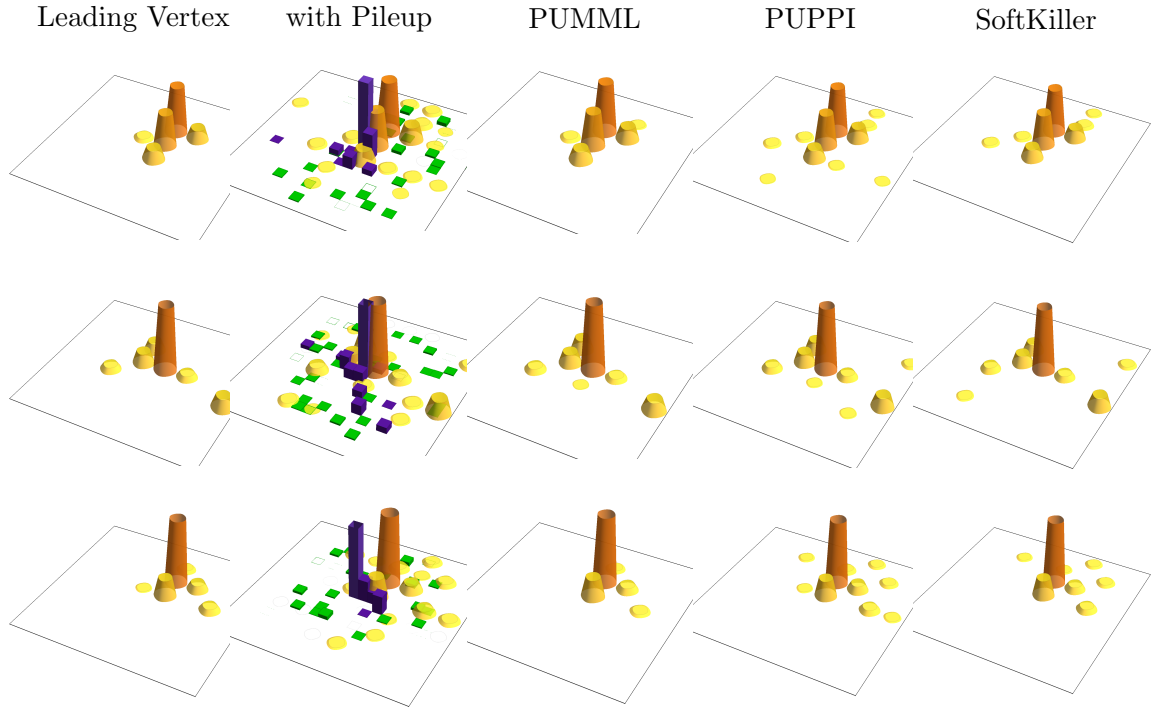

Samples were generated with the number of pileup vertices ranging from 0 to 180. Since the model must be trained to fix its parameters, the learned model depends on the pileup distribution used for training. For our pileup simulations, we trained on a Poisson distribution of NPUs with mean $\langle \text{NPU}\rangle = 140$. For robustness studies, we also tried training with NPU$=140$ for each event or NPU$=20$ for each event. The average jet image inputs for this sample are shown in Fig.~\ref{fig:jetimagesRGB}. For comparison, we show the performance of two powerful and widely used constituent-based pileup mitigation methods: PUPPI~\cite{Bertolini:2014bba} and SoftKiller~\cite{Cacciari:2014gra}. In both cases, default parameter values were used: $R_0=0.3$, $R_\text{min}=0.02$, $w_\text{cut} = 0.1$, $p_\text{T}^\text{cut}(\text{NPU}) = 0.1+0.007\times\text{NPU}$ (PUPPI), grid size = 0.4 (SoftKiller). Variations in the PUPPI parameters did not yield a large difference in performance. Both PUPPI and SoftKiller were implemented at the particle level and then discretized for comparison with PUMML. We show the action of the various pileup mitigation methods on a random selection of events in Fig.~\ref{fig:ex_imgs1}. On these examples, PUMML more effectively removes moderately soft energy deposits that are retained by PUPPI and SoftKiller.

\begin{figure}[t]
\centering
\includegraphics[scale = 0.58]{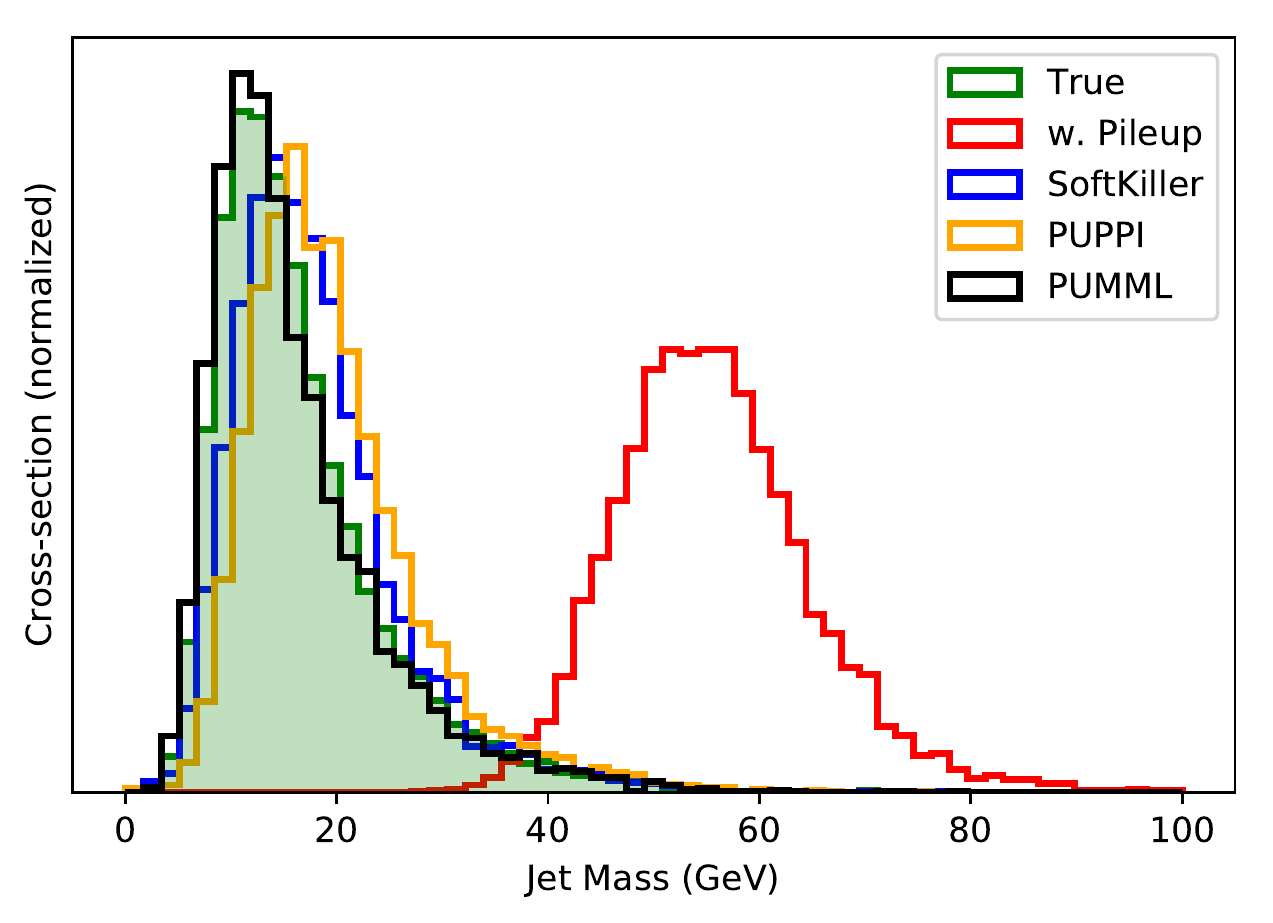}
\includegraphics[scale = 0.58]{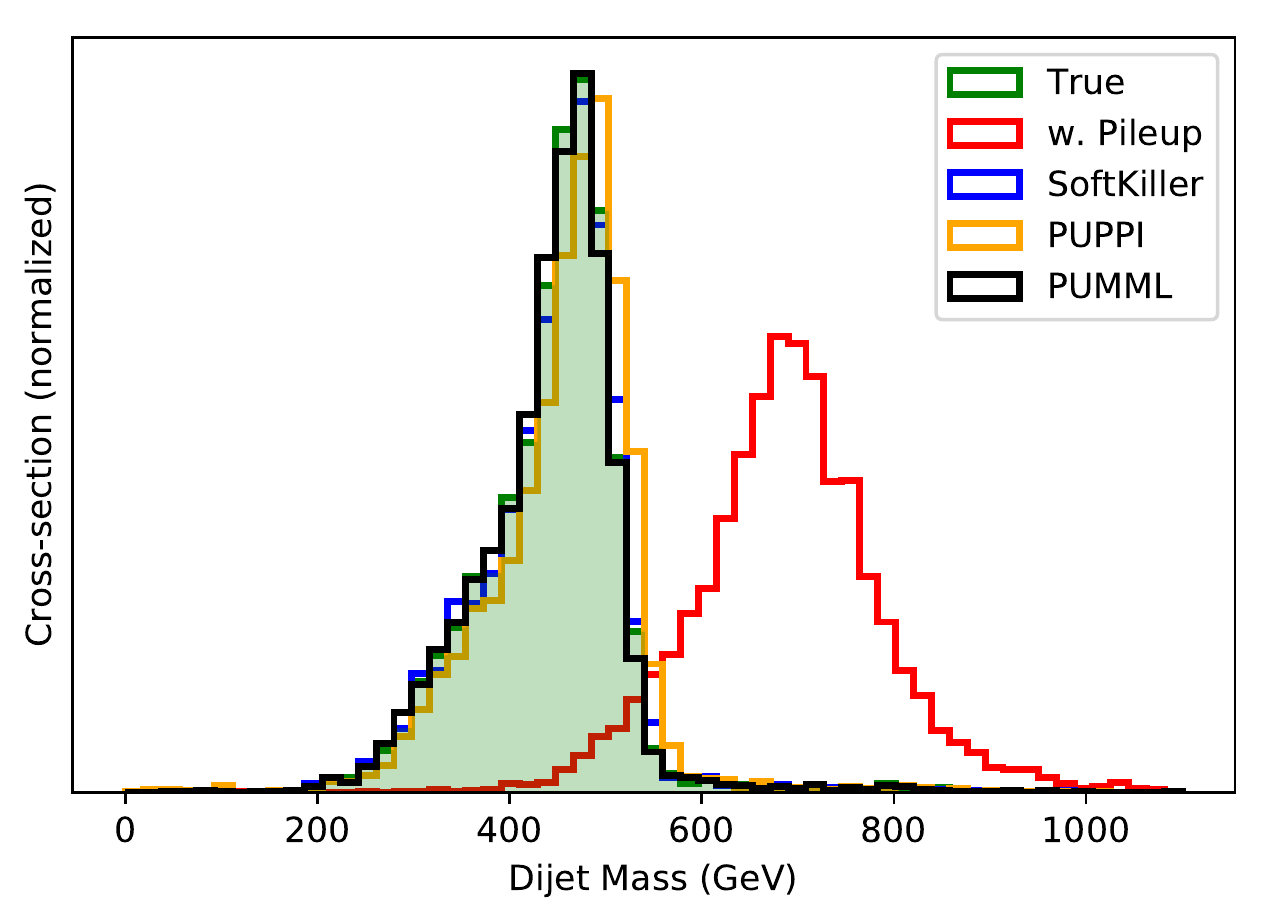}
\includegraphics[scale = 0.58]{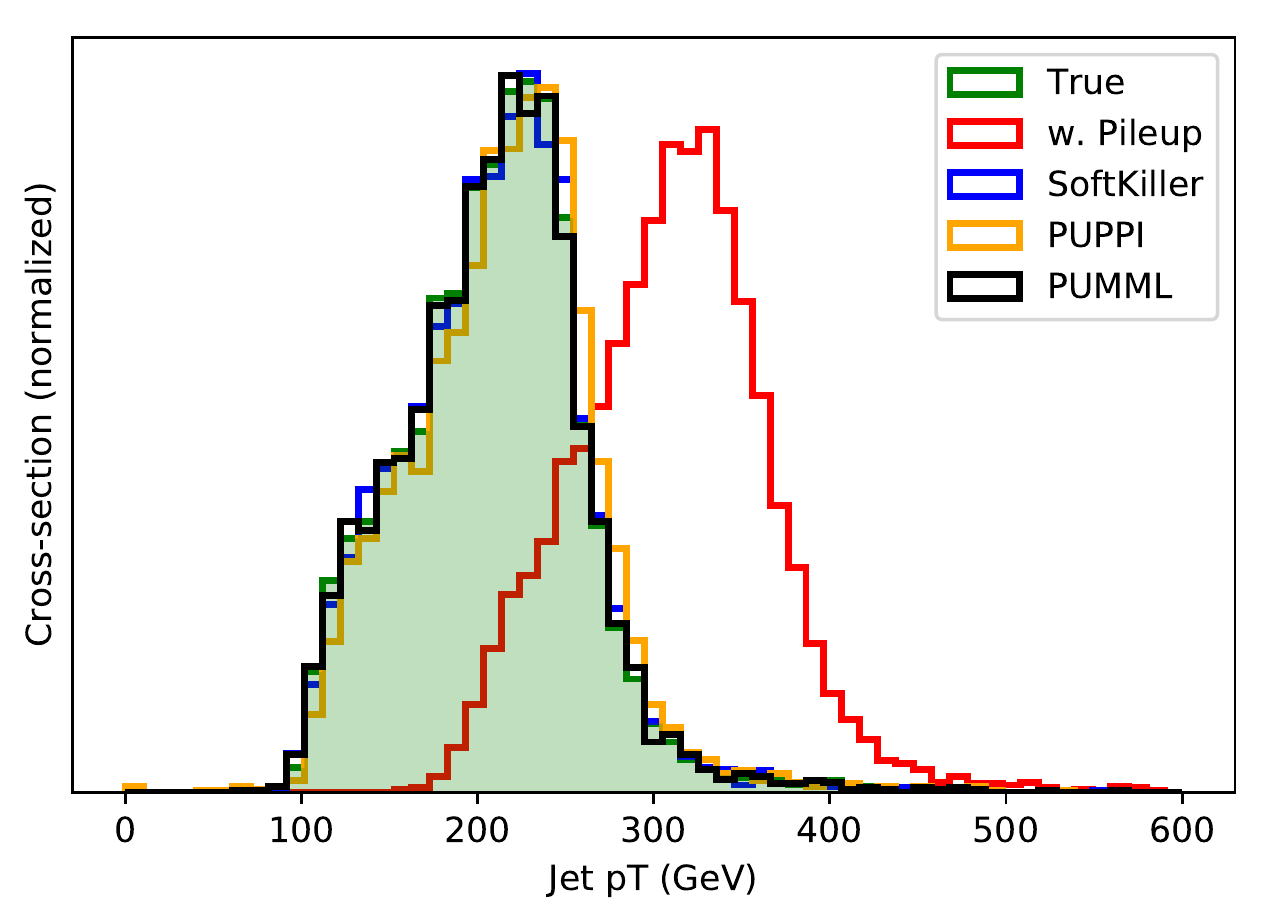}
\includegraphics[scale = 0.58]{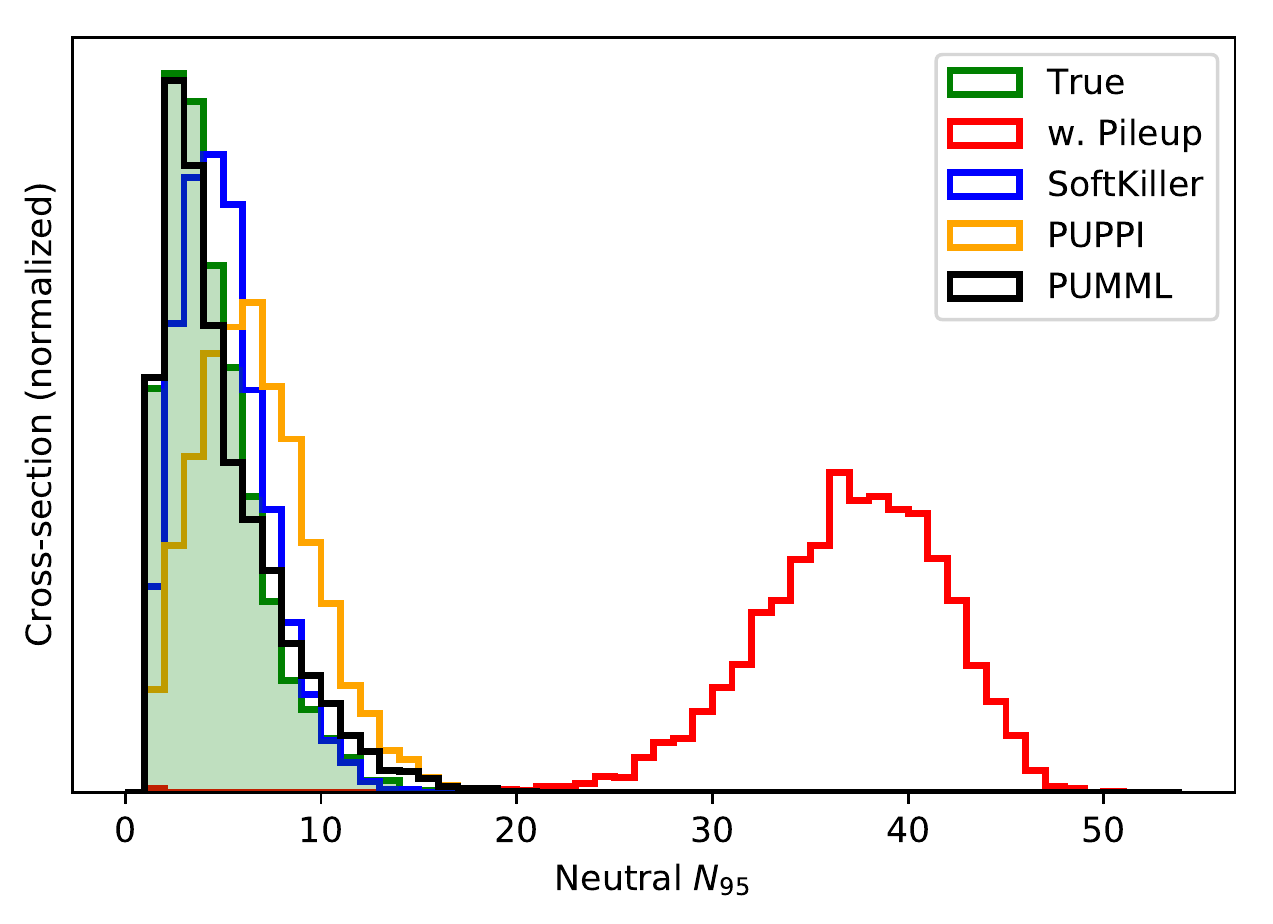}
\includegraphics[scale = 0.58]{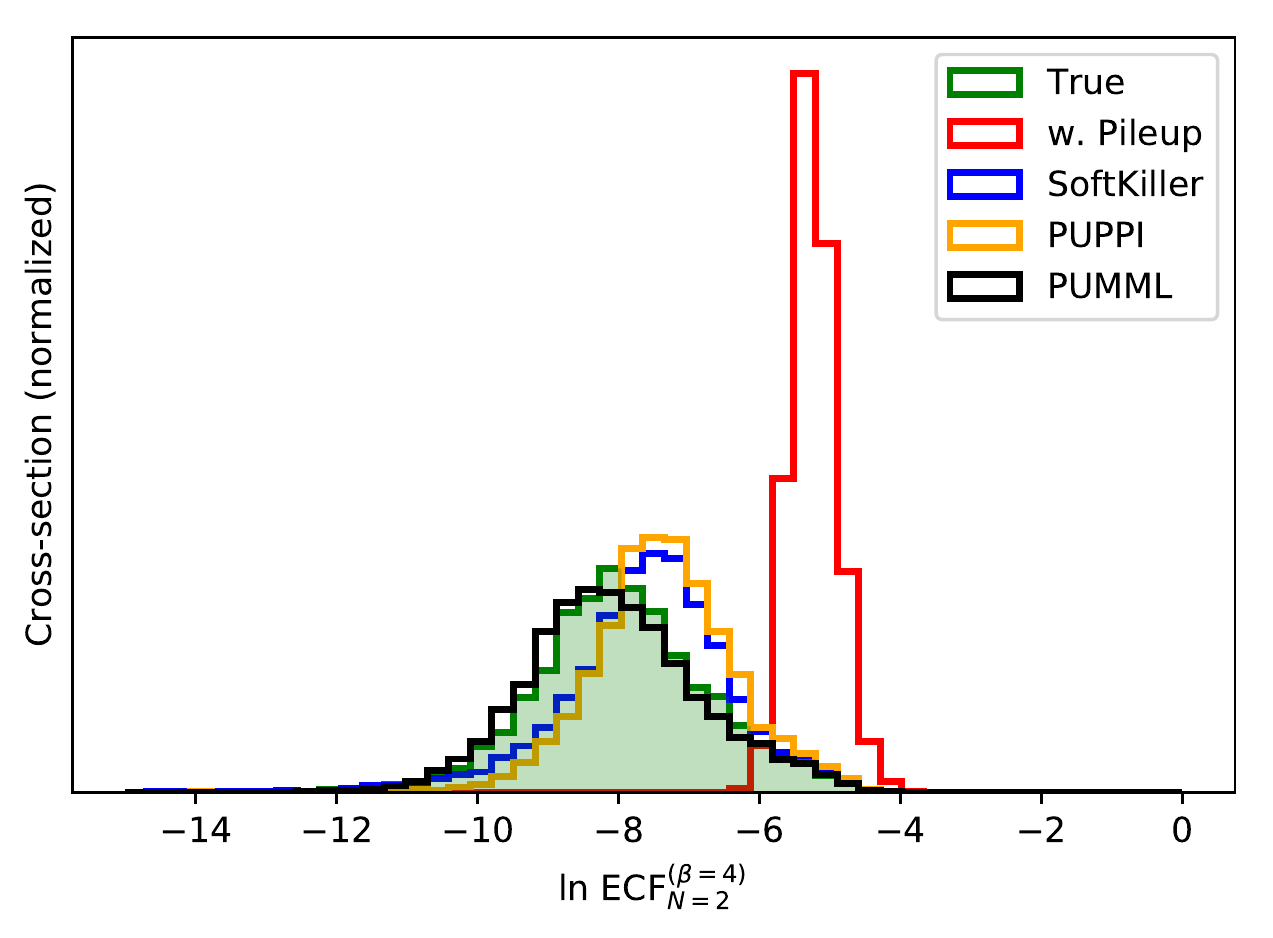}
\includegraphics[scale = 0.58]{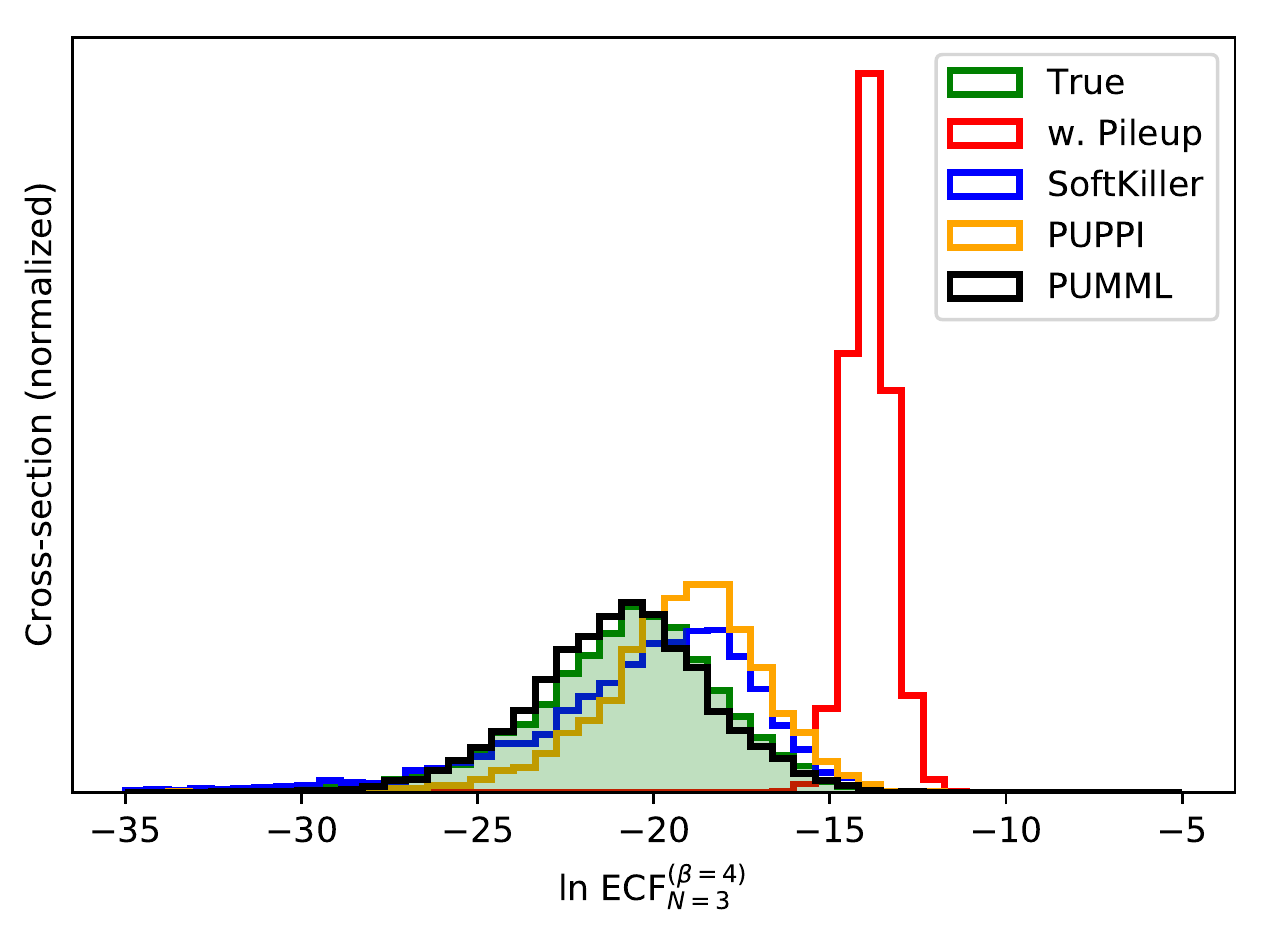}
\caption{\label{fig:dists} Distributions of leading jet mass (top left), dijet mass (top right), leading jet $p_T$ (middle left), neutral $N_{95}$ (middle right), $\ln$ ECF$_{N=2}^{(\beta = 4)}$ (bottom left), and $\ln$ ECF$_{N=3}^{(\beta = 4)}$ (bottom right) for the considered pileup subtraction methods with Poissonian $\langle \text{NPU} \rangle = 140$ pileup. While all of the pileup mitigation methods do well for observables such as the dijet mass and jet $p_T$, PUMML more closely matches the true distributions of more sensitive substructure observables like mass, neutral $N_{95}$, and the energy correlation functions.}
\end{figure}

\begin{figure}[t]
\centering
\includegraphics[scale = 0.58]{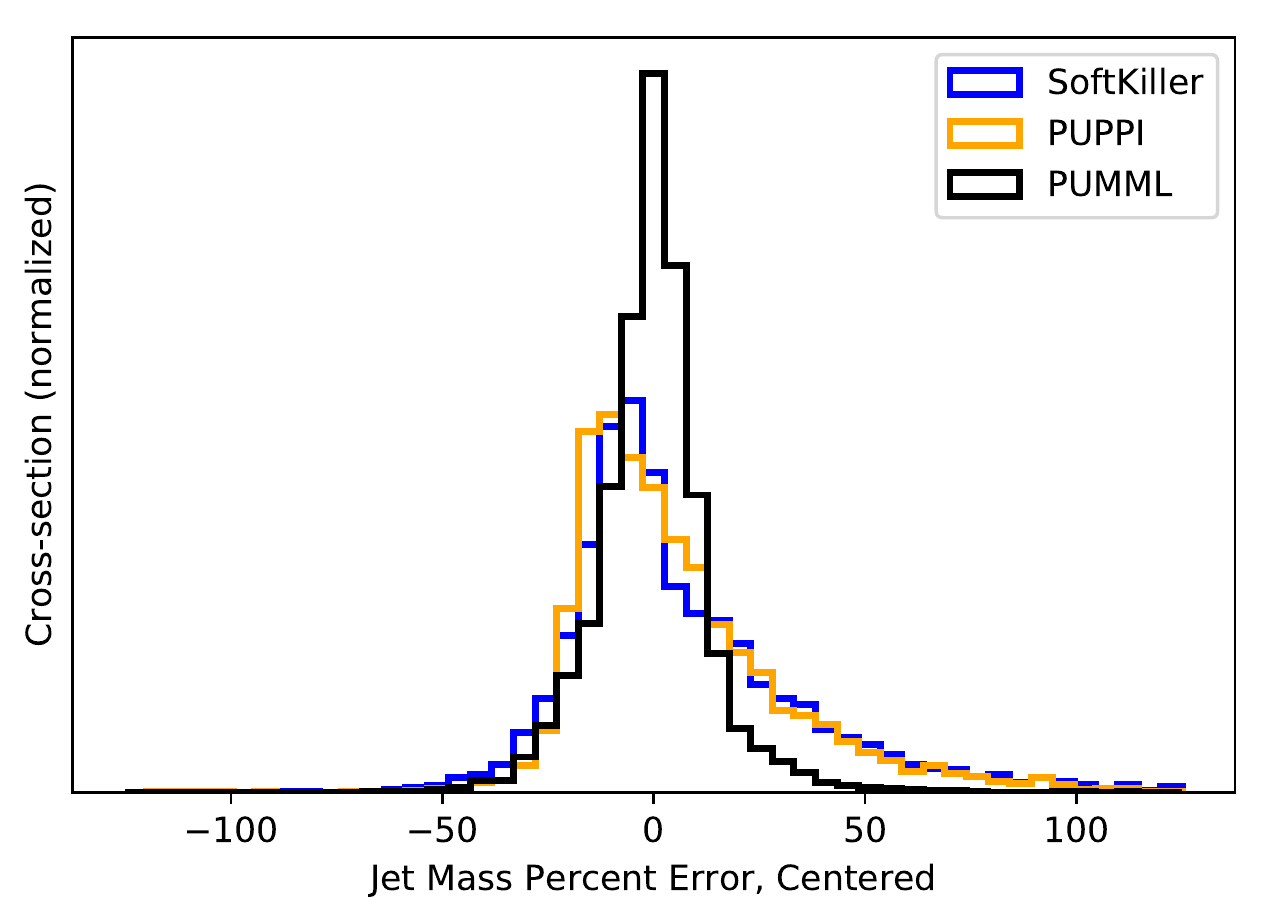}
\includegraphics[scale = 0.58]{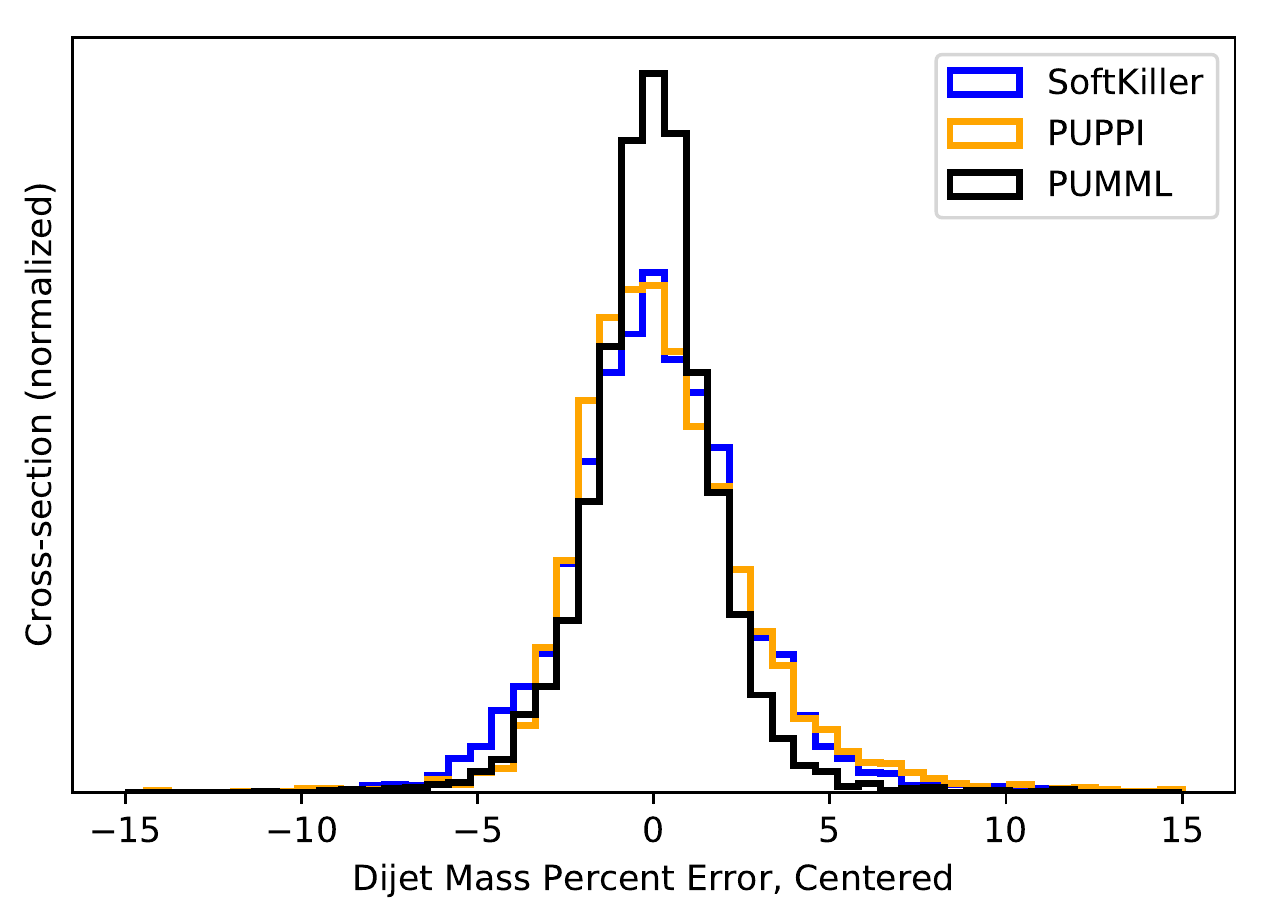}
\includegraphics[scale = 0.58]{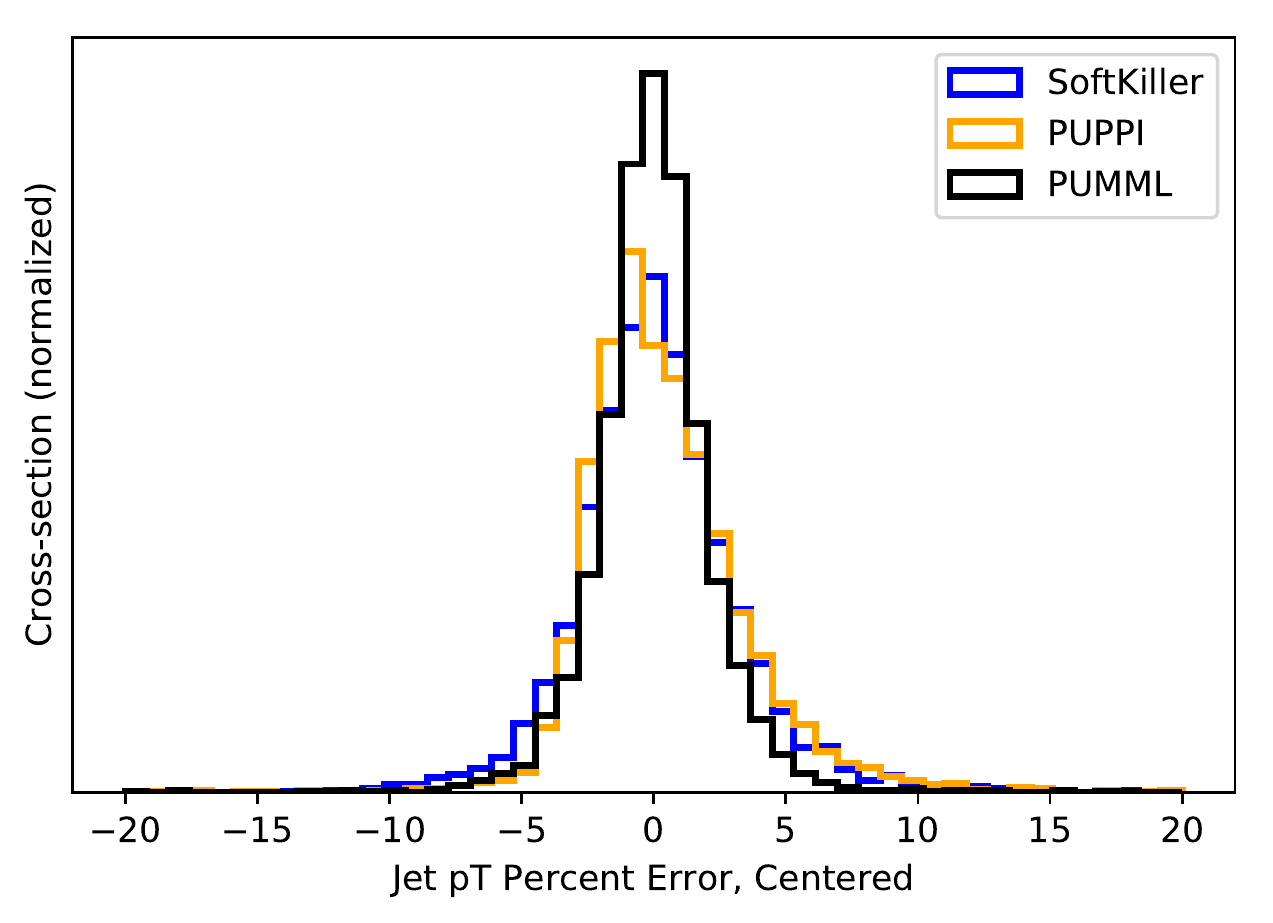}
\includegraphics[scale = 0.58]{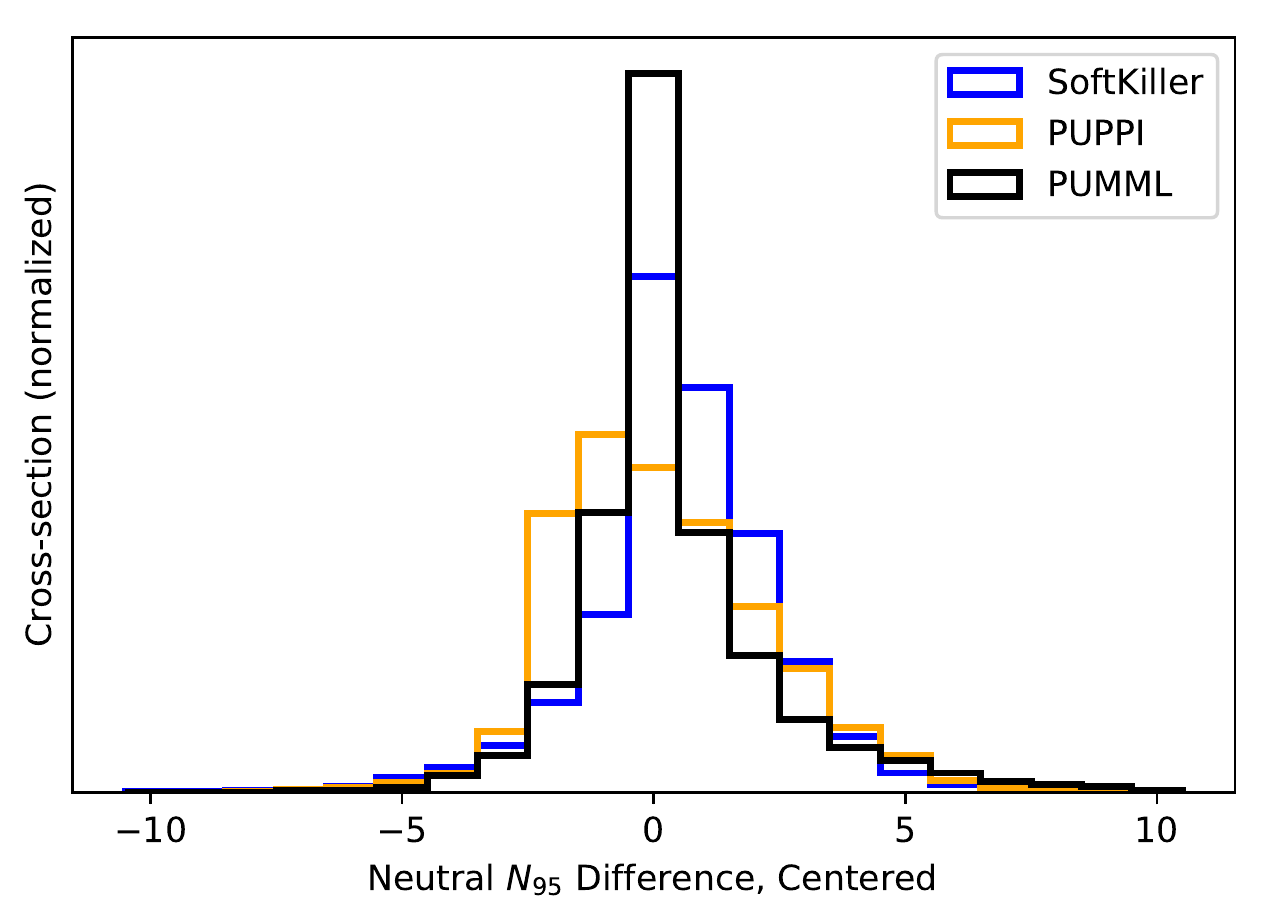}
\includegraphics[scale = 0.58]{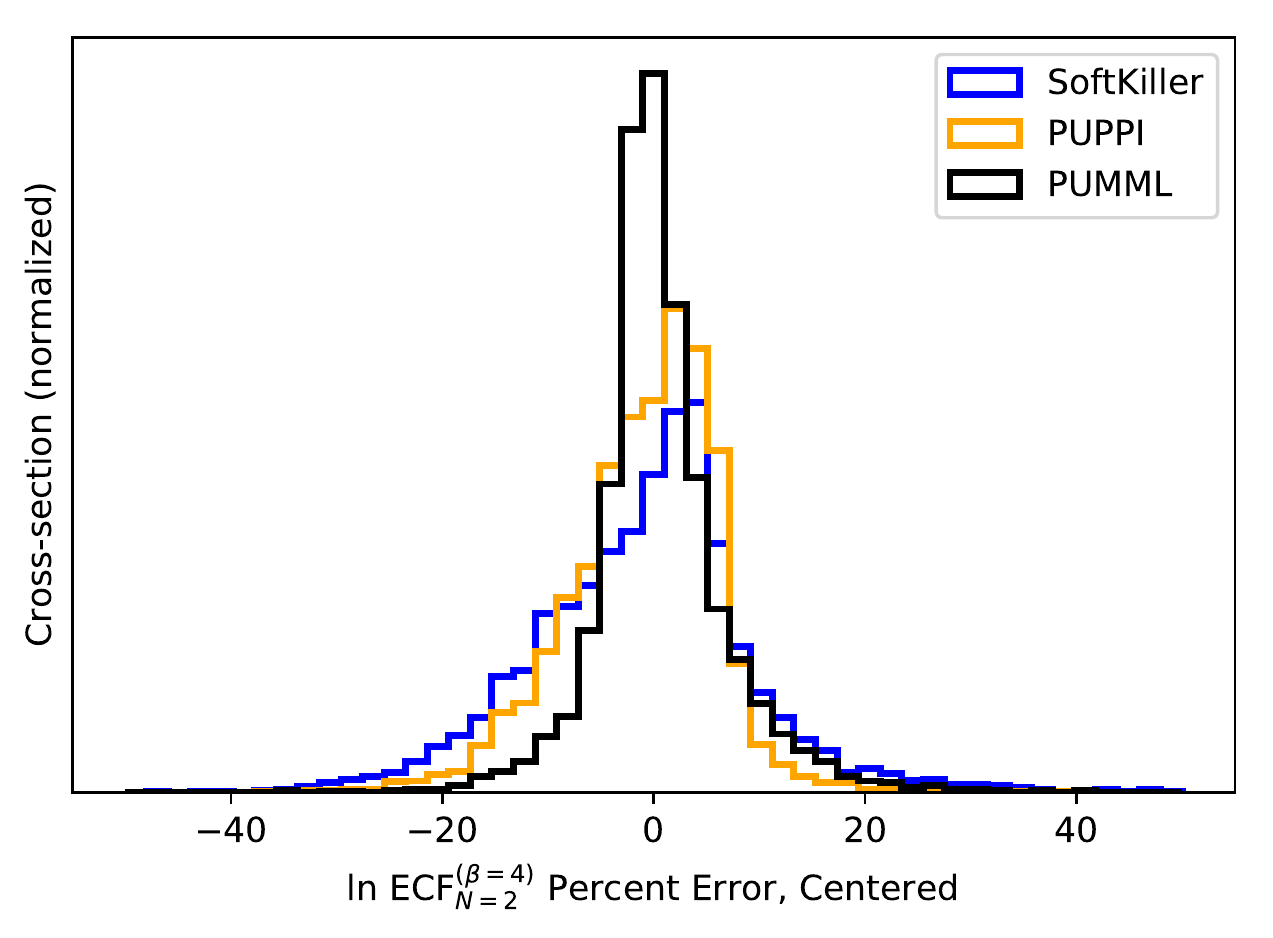}
\includegraphics[scale = 0.58]{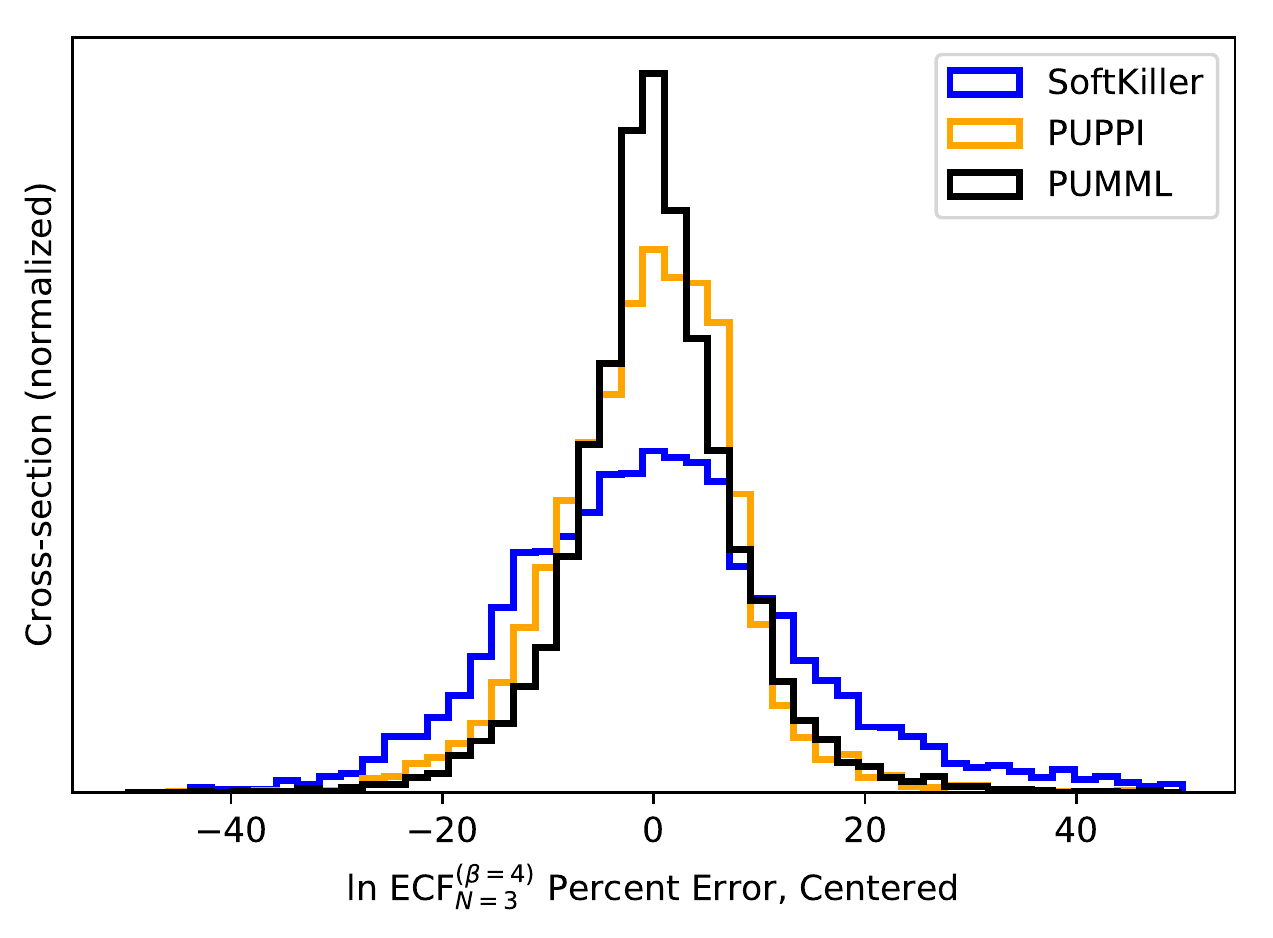}
\caption{\label{fig:diffhists}
Distributions of the percent error between reconstructed and true values for leading jet mass (top left), dijet mass (top right), leading jet $p_T$ (middle left), neutral $N_{95}$ (middle right), $\ln$ ECF$_{N=2}^{(\beta = 4)}$ (bottom left), and $\ln$ ECF$_{N=3}^{(\beta = 4)}$ (bottom right) for the considered pileup subtraction methods with Poissonian $\langle \text{NPU} \rangle = 140$ pileup. For the discrete neutral $N_{95}$ observable, only the difference is shown. All distributions are centered to have median at 0. The improved reconstruction performance of PUMML is highlighted by its taller and narrower peaks.}
\end{figure}

To evaluate the performance of different pileup mitigation techniques, we compute several observables and compare the true values to the corrected values of the observables. To facilitate a comparison with PUMML, which outputs corrected neutral calorimeter cells rather than lists of particles, a detector discretization is applied to the true and reconstructed events. Our comparisons focus on the following six jet observables:
\begin{itemize}
\item \emph{Jet Mass}:  Invariant mass of the leading jet.
\item \emph{Dijet Mass}: Invariant mass of the two leading jets.
\item \emph{Jet Transverse Momentum}: The total transverse momentum of the jet.
\item \emph{Neutral Image Activity}, $N_{95}$~\cite{Pumplin:1991kc}: The number of neutral calorimeter cells which account for 95\% of the total neutral transverse momentum.
\item \emph{Energy Correlation Functions}, ECF$_N^{(\beta)}$~\cite{larkoski2013}: Specifically, we consider the logarithm of the two- and three-point ECFs with $\beta = 4$.
\end{itemize}

Fig.~\ref{fig:dists} illustrates the distributions of several of these jet observables after applying the different pileup subtraction methods. While these plots are standard, they do not give a per-event indication of performance. A more useful comparison is to show the distributions of the per-event percent error in reconstructing the true values of the observables, which are shown in Fig.~\ref{fig:diffhists}. To numerically explore the event-by-event effectiveness, we can look at the Pearson linear correlation coefficient between the true and corrected values or the interquartile range (IQR) of the percent errors. Table~\ref{tab:corrs} summarizes the event-by-event correlation coefficients of the distributions shown in Fig.~\ref{fig:dists}. Table~\ref{tab:iqr} summarizes the IQRs of the distributions shown in Fig.~\ref{fig:diffhists}. PUMML outperforms the other pileup mitigation techniques on both of these metrics, with  improvements for jet substructure observables such as the jet mass and the energy correlation functions.

\begin{table}[t]
\centering
\begin{tabular}{|l|c|c|c|c|c|}
\hline
\bf Correlation (\%) & w. Pileup & PUMML & PUPPI & SoftKiller \\
\hline
Jet mass & 65.5 & 97.4 & 94.0 & 91.3\\
Dijet mass & 85.5& 99.5&  95.8 & 99.1 \\
Jet $p_T$ & 94.4& 99.7 & 98.0 & 99.4\\
Neutral $N_{95}$ & 36.2 & 75.3 & 70.4 & 67.7\\
$\ln$ ECF$_{N=2}^{(\beta = 4)}$ &  60.4  & 90.5 &  83.3  & 68.8 \\
$\ln$ ECF$_{N=3}^{(\beta = 4)}$ & 41.6 & 77.2 & 69.1 & 45.7\\
\hline
\end{tabular}
\caption{\label{tab:corrs} Correlation coefficients between the true and corrected values of different jet observables on an event-by-event level. The first column lists the correlation without any pileup mitigation applied to the event. Larger correlation coefficients are better.\\}
\end{table}

\begin{table}[t]
\centering
\begin{tabular}{|l|c|c|c|c|c|}
\hline
\bf IQR (\%) &  PUMML & PUPPI & SoftKiller \\
\hline
Jet mass & 13.0 & 28.7 & 30.8 \\
Dijet mass & 2.02 &  2.95 & 2.97 \\
Jet $p_T$ & 2.26 & 3.40 & 3.39\\
$\ln$ ECF$_{N=2}^{(\beta = 4)}$ & 5.63 & 8.82 & 11.9  \\
$\ln$ ECF$_{N=3}^{(\beta = 4)}$ & 8.48 & 10.7 & 16.7 \\
\hline
\end{tabular}
\caption{\label{tab:iqr} 
The interquartile ranges (IQR) of the distributions in Fig.~\ref{fig:diffhists}. Note that PUMML performs better than either PUPPI or SoftKiller. Lower IQR indicates better performance. 
}
\end{table}

\section{Robustness}
\begin{figure}[t]
\centering
\includegraphics[scale=.85]{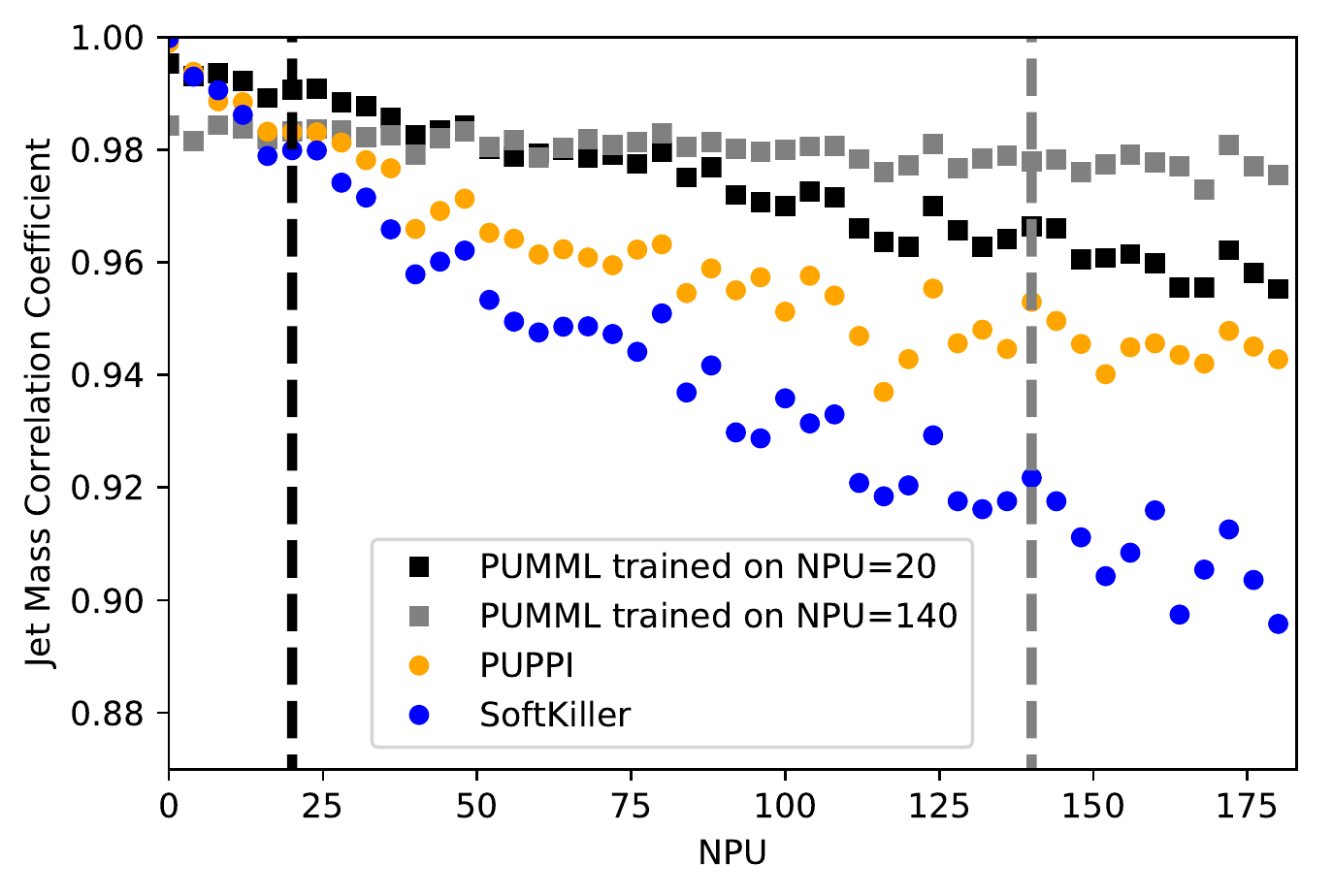}
\caption{\label{fig:rob}Correlation coefficients between reconstructed and true jet masses plotted as a function of NPU for the different pileup mitigation schemes. PUMML was trained on 50k events with either NPU$=20$ or NPU$=140$ indicated by dashed vertical lines.
The performance of PUMML with Poissonian $\langle \text{NPU} \rangle = 140$ is similar to the NPU$=140$ curve.  PUMML is surprisingly performant well outside the NPU range on which it was trained and consistently outperforms PUPPI and SoftKiller. Note that PUMML trained on the lower NPU sample better reconstructs the jet mass in the low pileup regime.}
\end{figure}

It is important to verify that PUMML learns a pileup mitigation function which is not overly sensitive to the NPU distribution of its training sample. Robustness to the NPU on which it is trained would indicate that PUMML is learning a universal subtraction strategy. To evaluate this robustness, PUMML was trained on 50k events with either $\text{NPU}=20$ or $\text{NPU}=140$ and then tested on samples with different NPUs. Fig.~\ref{fig:rob} shows the jet mass correlation coefficients as a function of the test sample NPU. PUMML learns a strategy that is surprisingly performant outside of the NPU range on which it was trained. Further, we see that by this measure of performance, PUMML consistently outperforms both PUPPI and SoftKiller.

\begin{figure}[t]
\centering
\includegraphics[scale=.85]{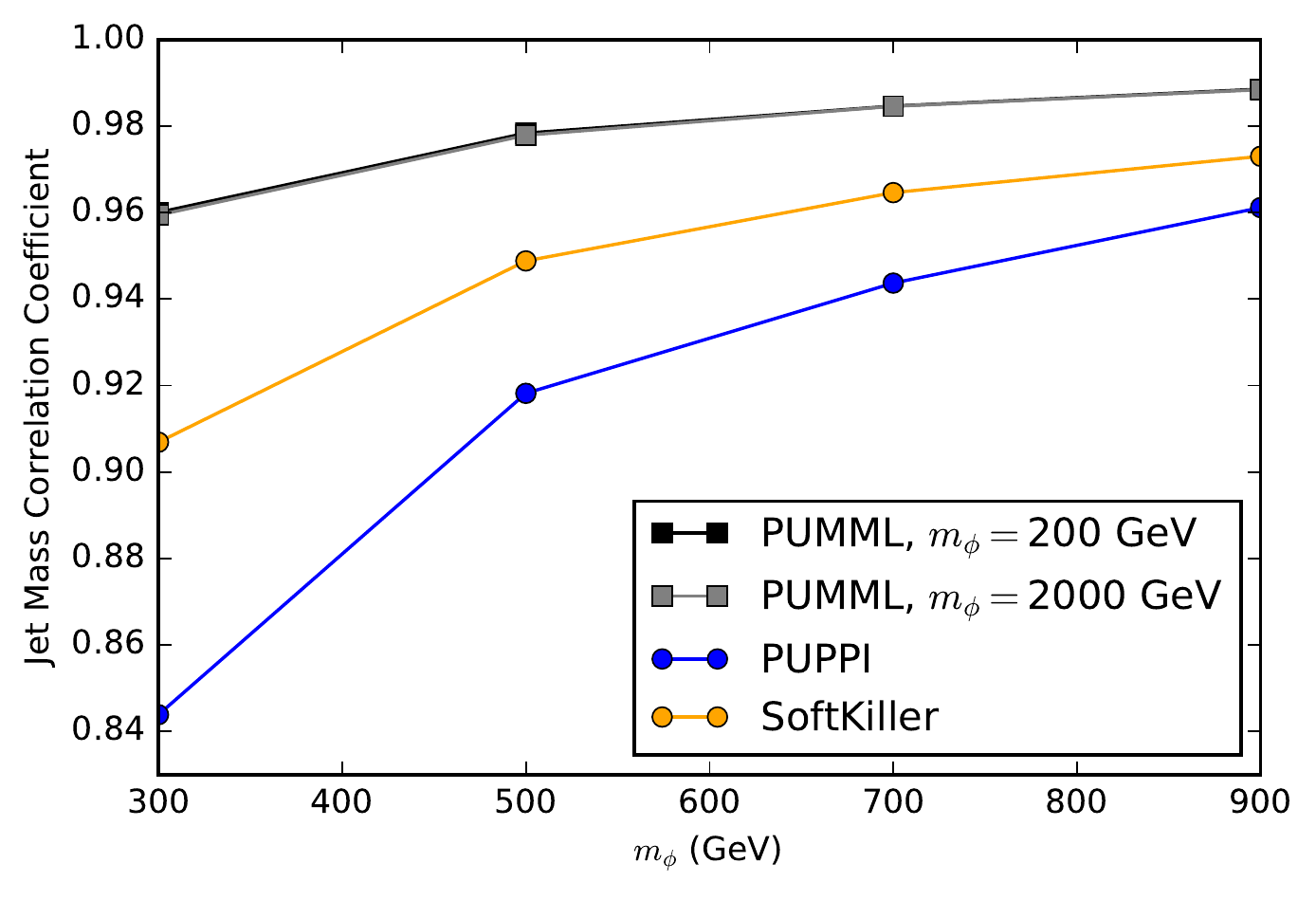}
\caption{\label{fig:pTsweep}Correlation coefficients between reconstructed and true jet masses plotted as a function of the mass of the scalar resonance with NPU=140.  A spread in scalar resonances is generated in order to produce a range in jet transverse momenta. In order to assess the impact of the $p_\text{T}$ distribution used for training, one version of PUMML was trained with a scalar mass of 200 GeV (black) and one was trained with a mass of 2 TeV (gray).  The two PUMML curves closely match one another.}
\end{figure}

A related robustness test is to probe how the performance of PUMML depends on the  $p_T$ spectrum of the training sample.
To explore this, we generated two large training samples (50k events): one with a scalar mass of 200 GeV and one with a scalar mass of 2 TeV; we did not impose any parton-level $p_T$ cuts on these samples.  After training these two networks, we tested them on a set of samples generated from scalars with intermediate masses, from 300 GeV to 900 GeV. As can be seen in Fig.~\ref{fig:pTsweep}, the performance of PUMML is very robust to the $p_T$ distribution of the jets in the training sample: the networks trained on the 200 GeV resonance and the 2 TeV resonance have identical performance. The figure also shows that the performance of PUMML is less sensitive to of the $p_T$ of the testing sample than either  PUPPI or Soft-Killer. This robustness test speaks to the PUMML algorithm's ability to learn universal aspects of pileup mitigation. 

A number of modifications of PUMML were also tried. Locally connected layers were tried instead of convolutional layers and were found to perform worse due to a large increase in the number of parameters of the model, while losing the translation invariance that makes PUMML powerful. We tried training without various combinations of the input channels; the model was found to perform moderately worse without either of the charged channels but suffered severe degradation without the total neutral channel. We tried using simpler models with only one layer or fewer filters per layer. Remarkably, even with only a single layer and a single $4\times 4$ filter (a model that has just 49 parameters), PUMML performed only moderately worse than the version presented in this study, which was allowed to be more complicated in order to achieve even better performance. 

\newpage
\section{What is PUMML learning?}
\begin{figure}[t]
\centering
\includegraphics[scale = 0.5]{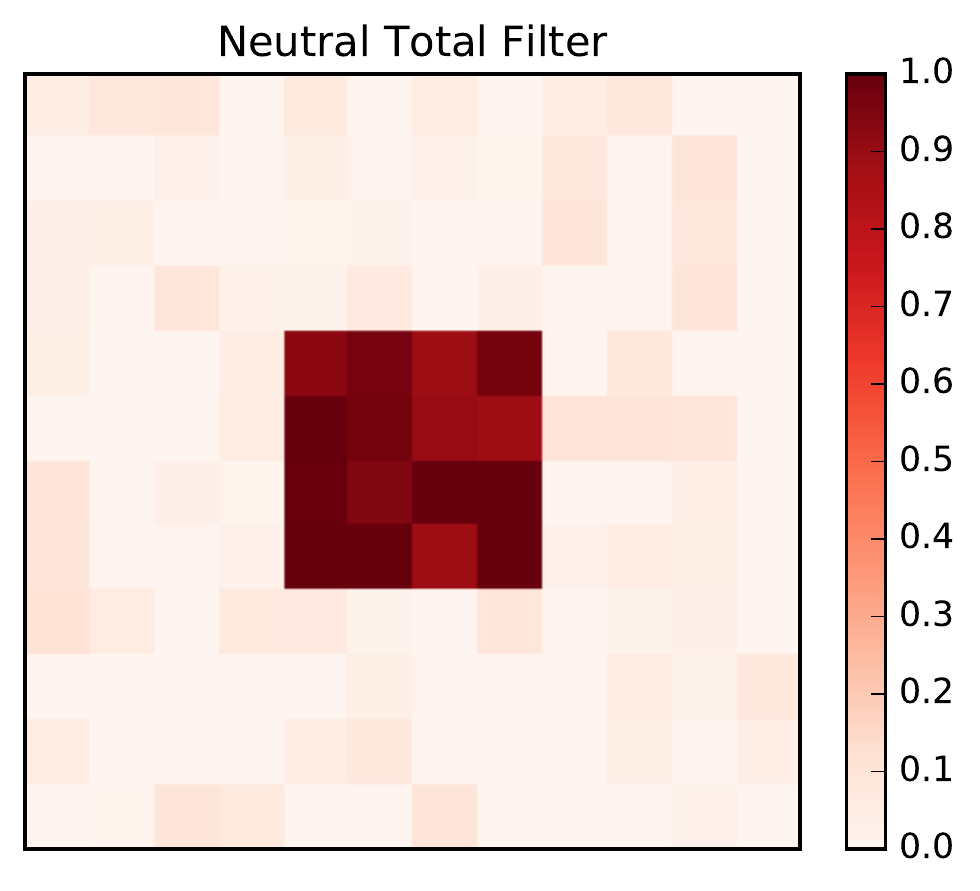}
\includegraphics[scale = 0.5]{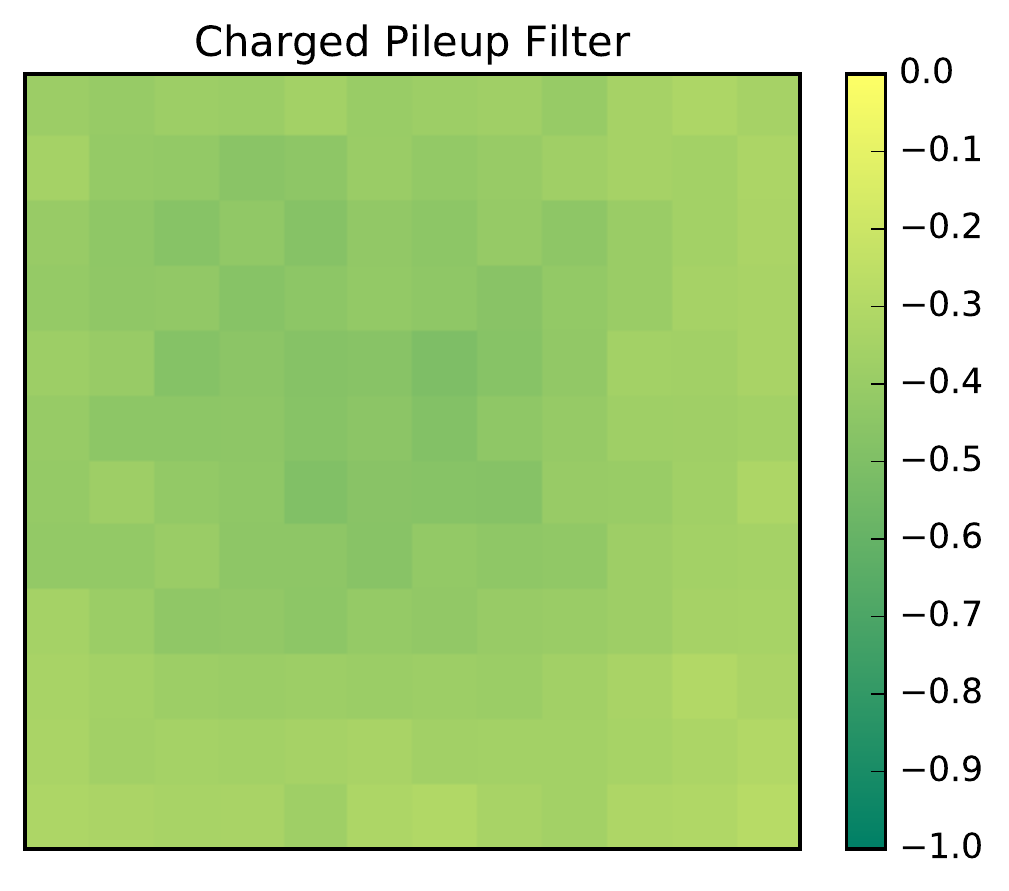}
\includegraphics[scale = 0.5]{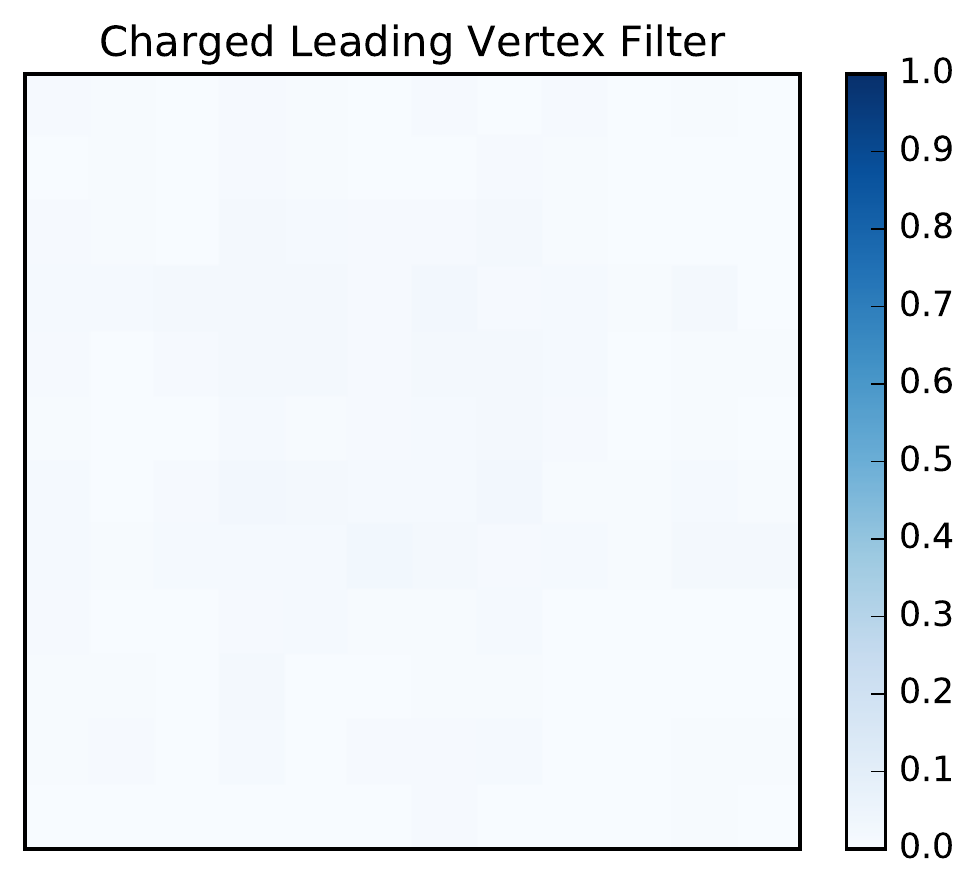}
\caption{\label{fig:filters} Filter weights for a simple PUMML network with a single $12\times 12$ filter and a ReLU activation function trained with $\langle\text{NPU}\rangle = 140$. The network has selected the relevant neutral pixel, turned off the charged leading vertex contribution, and is using the charged pileup information uniformly.}
\end{figure}

While it is generally very difficult to determine what a network is learning, one possible probe is to examine the weights of the filter layers in the convolutional network. For our full network, these weights are complicated and the subtractor that the network learns is difficult to probe analytically. Instead, we trained a simplified PUMML network with a single $12\times12$ pixel filter, which spans $3\times 3$ neutral pixels, with no bias term. The different channels of this filter are shown in Fig.~\ref{fig:filters}. The neutral filter clearly selects the relevant neutral pixel for subtraction, while the charged pileup filter is approximately uniform (with the value dependent on the specific choice of loss function and activation function), and the charged leading vertex filter does not significantly contribute. 

The filter values motivate the following parameterization of what PUMML is learning:
\begin{equation}\label{eq:smallsub}
p_T^{\text{N,LV}} = 1.0 \,p_T^{\text{N,total}} - \beta \, p_T^{\text{C,PU}} + 0.0 \, p_T^{\text{C,LV}} ,
\end{equation}
for some $\mathcal O(1)$  constant $\beta$, where $p_T^{\text{N,LV}}$, $p_T^{\text{N,total}}$,  $p_T^{\text{C,PU}}$, and $ p_T^{\text{C,LV}}$ are the neutral-pixel-level transverse momenta of the neutral leading-vertex particles, all neutral particles, charged pileup particles, and charged leading-vertex particles, respectively. The values 1.0 and 0.0 in Eq.~\eqref{eq:smallsub} are stable (to the 0.05 level) under variations in the loss and activation functions. This is reassuring as the learned subtractor is thereby robust in the NPU $\to 0$ limit despite begin trained on $\langle \text{NPU} \rangle = 140$.

Eq.~\eqref{eq:smallsub} is remarkably similar to the physically-motivated formula used in Jet Cleansing~\cite{Krohn:2013lba}. Cleansing is built on the observation that since pileup is the incoherent sum of many separate scattering events, its variance is smaller than the variance of the radiation from the leading-vertex. Thus, it is better to estimate $p_T^{\text{N,PU}}$ from $p_T^{\text{C,PU}}$ than to estimate $p_T^{\text{N,LV}}$ from $p_T^{\text{C,LV}}$. The simplest form of Cleansing (Linear Cleansing) gives the formula:
\begin{equation}
p_T^{\text{N,LV}} =  p_T^{\text{N,tot}} -\left(\frac{1}{\overline{\gamma_0}}-1\right)\, p_T^{\text{C,PU}},
\end{equation}
where $\overline{\gamma_0}$ is the average ratio of charged $p_T$ to total $p_T$ in a subjet. Thus this simple one $12\times12$ filter PUMML network is learning a subtractor of precisely the same parametric form as Linear Cleansing!

The value of $\overline{\gamma_0}$ in Linear Cleansing and the value of  $\beta$ that is learned in Eq.~\eqref{eq:smallsub} depend on how soft radiation is handled. For example, if no reconstruction threshold is applied, $\overline{\gamma_0} \approx 2/3$ (since 2/3 of pions are charged). In addition, the value of $\beta$ depends on the loss function used. For example, if the loss function is minimized when the means of the true and predicted neutral transverse momenta are equal:
\begin{equation}\label{eq:meanabs}
\ell=\left|  \langle p_T^{\text{(true)}}\rangle  -  \langle p_T^{\text{(pred)}} \rangle
\right|
=\left|
  \langle p_T^{\text{N,LV}} \rangle
  - \langle p_T^{\text{N,total}}\rangle 
  + \beta \langle\, p_T^{\text{C,PU}}\rangle 
  \right|,
\end{equation}
then we find that the optimal $\beta$ is:
\begin{equation}
 \beta = \frac{\langle p_T^{\text{N,PU}}\rangle}{\langle p_T^{\text{C,PU}} \rangle}.
 \label{betalin}
 \end{equation}
Training the $12\times 12$ PUMML filter without a ReLU or bias term, using the loss function of Eq.~\eqref{eq:meanabs} with the average taken pixel-wise over the batch, we find $\beta = 0.59$ with no charged reconstruction cut and $\beta = 1.18$ with the cut. These values are consistent with those predicted by Eq.~\eqref{betalin} of 0.62 and 1.26, respectively.

On the other hand, if we take a mean squared error loss function:
\begin{equation}
\ell=\left\langle \left(p_T^{\text{(true)}}  -  p_T^{\text{(pred)}}\right)^2 \right\rangle,
\end{equation}
then the minimum occurs at:
\begin{equation}
\beta = \frac{ \langle p_T^{\text{N,PU}}p_T^{\text{C,PU}} \rangle }{ \langle p_T^{\text{C,PU}} 
p_T^{\text{C,PU}} \rangle }.
\label{betaquad}
\end{equation}
This still depends only on the pileup properties, as with Linear Cleansing, but also depends on correlations between neutral and charged radiation. For example, training the $12\times12$ PUMML filter without a ReLU or bias term using a mean squared error loss function, we find $\beta=0.56$ with no charged reconstruction cut and $\beta =0.97$ with the cut. These numbers are in general agreement (within $10-20$\%) with a direct evaluation of the right-hand side of Eq.~\eqref{betaquad}. In the limit that neutral and charged pileup radiation are constant, Eq.~\eqref{betaquad} reduces to Eq.~\eqref{betalin}.

Whether the loss function of Eq.~\eqref{betalin} or Eq.~\eqref{betaquad} (or something else entirely) is better is not simple to establish. The inclusion of the ReLU activation function further complicates the analysis since the model is equally penalized for all non-positive predictions. We find with the single $12\times 12$ filter, using the loss function of Eq.~\eqref{eq:loss} and including a ReLU and bias term, PUMML achieves a jet mass correlation coefficient of 90.4\%. This is competitive with the values listed in Table.~\ref{tab:corrs}, as we might expect since Linear Cleansing has comparable performance to PUPPI and SoftKiller. The full network improves on Linear Cleansing by exploiting additional correlations that are hard to disentangle by looking at the filters.

\section{Conclusions\label{sec:conc}}
In this paper, we have introduced the first application of machine learning to the critically important problem of pileup mitigation at hadron colliders. We have phrased the problem of pileup mitigation in the language of a machine learning regression problem. The method we introduced, PUMML, takes as input the transverse momentum distribution of charged leading-vertex, charged pileup, and all neutral particles, and outputs the corrected leading vertex neutral energy distribution. We demonstrated that PUMML works at least as well as, and often better than, the competing algorithms PUPPI and SoftKiller in their default implementations.  It will be exciting to see these algorithms compared with a full detector simulation, where it will be possible to test the sensitivity to important experimental effects such as resolutions and inefficiencies.

There are several extensions and additional applications of the PUMML framework beyond the scope of this study. As mentioned in Section~\ref{sec:algorithm}, PUMML can very naturally be extended from jet images to entire events. Applying this event-level PUMML to the problem of missing transverse energy would be a natural next step.  While the filter sizes can be the same for the event and jet images, the network training will likely require modification.  Furthermore, the inhomogeneity of the detector response with $|\eta|$ will require attention.  Another potentially useful modification to PUMML would be to train to predict the neutral pileup $p_T$ rather than the neutral leading vertex $p_T$ in order to increase out-of-sample robustness of the learned pileup mitigation algorithm. Additionally, using larger-$R$ jets may be of interest, thereby necessitating a resizing of the local patch or other PUMML parameters, all of which is easily achieved.

An important consideration when using machine learning for particle physics applications is how the method can be used with data and whether or not the systematic uncertainties are under control. Unlike a purely physically-motivated algorithm, such as PUPPI or SoftKiller, machine learning runs the risk of being a ``black-box'' which can be difficult to understand. Nevertheless, machine learning is powerful, scaleable, and capable of complementing physical insight to solve complicated or otherwise intractable problems.

To prevent the model from learning simulation artifacts, it is preferable to train on actual data rather than simulation. In many machine learning applications in collider physics, obtaining truth-level training samples in data is a substantial challenge. To overcome this challenge in classification tasks,~\cite{dnrs2017} introduces an approach to train from impure samples using class proportion information. For PUMML and pileup mitigation more broadly, a more direct method to train on data is possible. To simulate pileup, we overlay soft QCD events on top of a hard scattering process, both generated with Pythia. Experimentally, there are large samples of minimum bias and zero-bias (i.e. randomly triggered) data. There are also samples of relatively pileup-free events from low luminosity runs. Thus we can construct high-pileup samples using purely data. This kind of {\it data overlay} approach, which has already been used by experimental groups in other contexts~\cite{marshall14,haas17}, could be perfect for training PUMML with data. Therefore, an implementation of ML-based pileup mitigation in an actual experimental setting could avoid mis-modeling artifacts during training, thus adding more robustness and power to this new tool.

\acknowledgments

The authors would like to thank Philip Harris, Francesco Rubbo, Ariel Schwartzman and Nhan Tran for stimulating conversations, in particular for suggesting some of the extensions mentioned in the conclusions. We would also like to thank Jesse Thaler for helpful discussions. PTK and EMM would like to thank the MIT Physics Department for its support. Computations in this paper were run on the Odyssey cluster supported by the FAS Division of Science, Research Computing Group at Harvard University. This work was supported by the Office of Science of the U.S.~Department of Energy (DOE) under contracts DE-AC02-05CH11231 and DE-SC0013607, the DOE Office of Nuclear Physics under contract DE-SC0011090, and the DOE Office of High Energy Physics under contract DE-SC0012567. Cloud computing resources were provided through a Microsoft Azure for Research award. Additional support was provided by the Harvard Data Science Initiative.

\bibliography{pumml}
\bibliographystyle{JHEP-2}

\end{document}